\documentclass[12pt,a4paper]{article}

\usepackage{amsmath,amssymb,bbm,slashed,cancel}
\usepackage{cite,tabularx}
\usepackage{subfigure}
\usepackage{graphicx,color}

\allowdisplaybreaks

\usepackage[height=8.85in,width=6.4in]{geometry}

\newcommand{\TeV}{\,\textrm{TeV}}
\newcommand{\GeV}{\,\textrm{GeV}}

\numberwithin{equation}{section} 
\def\beq#1\eeq{\begin{align}#1\end{align}}

\definecolor{BlueViolet}{rgb}{0.2, 0.00, 0.7}
\definecolor{Blue}{rgb}{0.15, 0.00, 0.9}
\usepackage[colorlinks=true,linkcolor=Blue,citecolor=Blue,urlcolor=BlueViolet]{hyperref}

\begin{document}
\begin{titlepage}

\begin{center}

\hfill KEK--TH--2284\\

\vskip .35in

{\large\bf
Revisiting electroweak radiative corrections to $b\to s\ell\ell$ in SMEFT
}

\vskip .4in

{
  Motoi Endo$^{\rm (a,b)}$,
  Satoshi Mishima$^{\rm (a)}$, and 
\vspace{.2cm}
  Daiki Ueda$^{\rm (c)}$
}

\vskip 0.25in

$^{\rm (a)}${\it
Theory Center, IPNS, KEK, Tsukuba, Ibaraki 305-0801, Japan}

\vskip 0.1in

$^{\rm (b)}${\it
The Graduate University of Advanced Studies (Sokendai),\\
Tsukuba, Ibaraki 305-0801, Japan}

\vskip 0.1in

$^{\rm (c)}${\it 
Department of Physics, Faculty of Science, University of Tokyo, \\Bunkyo-ku, Tokyo 113-0033, Japan}
 
\vskip 0.1in

\end{center}

\vskip .3in

\begin{abstract}

We revisit electroweak radiative corrections to Standard Model Effective Field Theory (SMEFT) operators which are relevant for the $B$-meson semileptonic decays.
The one-loop matching formulae onto the low-energy effective field theory are provided without imposing any flavor symmetry.
The on-shell conditions are applied especially in dealing with quark-flavor mixings. 
Also, the gauge independence is shown explicitly in the $R_\xi$ gauge. 

\end{abstract}
\end{titlepage}

\setcounter{page}{1}
\renewcommand{\thefootnote}{\#\arabic{footnote}}
\setcounter{footnote}{0}

\section{Introduction}
\label{sec:introduction}
Flavor-changing neutral currents are a class of the novel observables which are sensitive to physics beyond the standard model (SM). 
In particular, semileptonic $B$-meson decays have exhibited deviations of the experimental results from the SM predictions.
In addition to the update on the angular analysis of $B^0 \to K^{*0}\mu^+\mu^-$ \cite{Aaij:2020nrf}, the LHCb collaboration has recently reported a result of $B^+ \to K^{*+}\mu^+\mu^-$ \cite{Gerick:2020}, which strengthens the need for yet-not-known contributions \cite{Ciuchini:2020gvn}. 
Besides, the lepton-flavor universality tests of $B \to K^{(*)}\ell^+\ell^-$ $(\ell=e,\mu)$ \cite{Aaij:2019wad,Aaij:2017vbb} have supported motivations for new physics (NP) contributions\footnote{In Ref.~\cite{Gabrielli:2019uqu}, long-distance QED contributions, induced by the magnetic-dipole corrections at the one loop,
to the $b\to s \ell\ell$ transitions, and massless dark-photon contributions have been investigated.}. 

The absence of new particle discoveries at the LHC experiments implies that NP exists in high energy scale.
In determining the NP scale by based on the flavor anomalies, it is important to examine radiative corrections carefully.
In fact, the above $b \to s\ell\ell$ anomalies require the scale of $\Lambda_{\rm NP}/g_{\rm NP} \sim 30\TeV$ if the contributions are generated at the tree level \cite{Ciuchini:2020gvn}, while it could become $\mathcal{O}(1)\TeV$ in the case when there are no tree-level contributions and they arise by radiative corrections.

In this article, we revisit electroweak (EW) radiative corrections to the semileptonic decays at the one-loop level, paying attention to quark transitions due to the Cabibbo-Kobayashi-Maskawa (CKM) matrix. 
If the NP scale is higher than the EW symmetry breaking (EWSB) scale, NP contributions are encoded in higher dimensional operators in the Standard Model Effective Field Theory (SMEFT) by integrating out NP particles.
Then, the flavor structures of these operators are affected by $W$-boson loops due to the CKM matrix, especially when the operators are flavor conserving at the NP scale, {\it i.e.}, they do not change flavors initially. 

Such radiative corrections have been explored in Ref.~\cite{Aebischer:2015fzz}, where the result was presented for the operators involving the right-handed top quark.\footnote{Such contributions to $b\to s\ell\ell$ have been studied in Ref.~\cite{Camargo-Molina:2018cwu} by based on Ref.~\cite{Aebischer:2015fzz}.}
Those with the left-handed up-type quarks have been studied in Ref.~\cite{Hurth:2019ula}, though the analysis was restricted to the flavor-universal operators under the assumption of U(3)$^5$ flavor symmetry. 
On the other hand, Refs.~\cite{Coy:2019rfr,Aebischer:2020lsx,Alasfar:2020mne} has attempted to evaluate those effects by solving the renormalization group equations (RGEs) especially of the SM Yukawa matrices. 
In contrast, we study {\it general} SMEFT operators which are relevant for $b \to s\ell\ell$. 
We do not assume any flavor symmetry {\it a priori}, or not rely on the RGEs, but evaluate one-loop diagrams explicitly. 
Besides, we pay special attention to the renormalization of flavor mixings of the external quark fields in the effective operators, which is inevitable due to the mass splitting between the quark generations and has been explored originally in the EW theory of the SM \cite{Aoki:1982ed,Bohm:1986rj}.
Here, the on-shell renormalization condition is adopted.\footnote{
 In Ref.~\cite{Dekens:2019ept}, an alternative approach was used for the field redefinition in studying the one-loop matching conditions.
}
We also perform the calculations in the $R_{\xi}$ gauge and show the gauge independence of the results explicitly.
Finally, we will derive analytic formulae for the EW radiative corrections to $b \to s\ell\ell$ in the SMEFT at the one-loop level, and apply them to study the current $b \to s\ell\ell$ anomalies.

\section{Framework}

In this article, we calculate the EW radiative corrections to the SMEFT operators. 
In an effective field theory approach, UV theory is matched onto the SMEFT at the NP scale by integrating out heavy degrees of freedom concerning the NP particles.
Then, the SMEFT Wilson coefficients are evolved down by solving the RGEs \cite{Jenkins:2013zja,Jenkins:2013wua,Alonso:2013hga,Endo:future}.
At the EWSB scale, they are matched onto the low-energy effective field theory (LEFT) by integrating out the EW bosons $(W, Z, H)$ and the top quark $(t)$.
The $b \to s\ell\ell$ observables are represented in terms of the LEFT. 
The purpose of this article is to provide the EW one-loop matching formulae at the EWSB scale, paying attention to flavor transitions due to the CKM matrix. 

We assume that the operators in the SMEFT and LEFT are represented in the mass eigenstate basis. 
In particular, so-called ``the down basis'' is adopted for the quark doublet, where the down-type quarks are chosen to be flavor diagonal. 
Then, in the SMEFT, the quarks are expressed as
\begin{align}
 &
 q = (V^\dagger u_L, d_L)^T,~~~
 u_R,~~~
 d_R,
\end{align}
where $V$ is the CKM matrix.
Similarly, the leptons are shown by $\ell = (\nu_L, e_L)^T$ and $e_R$, where the neutrino masses and the Pontecorvo-Maki-Nakagawa-Sakata (PMNS) matrix are neglected. 
On the other hand, the ferimons in the LEFT are regarded as Dirac fields in the mass eigenstate basis. 

The SMEFT Lagrangian is represented as
\begin{align}
 \mathcal{L}_{\rm eff} = \mathcal{L}_{\textrm{SM}} + \sum_i C_i \mathcal{O}_i,
 \label{eq:SMEFT}
\end{align}
where the first term in the right-hand side is the SM Lagrangian, and the second term corresponds to the higher dimensional operators, for which we work with the Warsaw basis \cite{Grzadkowski:2010es}.
In this article, we consider the operators which are relevant for the current $b \to s\ell\ell$ anomalies.
Besides, we are interested in the matching conditions at the one-loop level. 
Since some of them have already been provided in Refs.~\cite{Aebischer:2015fzz,Hurth:2019ula}, we will not study them here. 
Rather, we focus on the case when the operators are flavor non-universal and affected by the $W$-boson loops. 
Then, the operators in interest are listed as
\begin{align}
 (\mathcal{O}^{(1)}_{\ell q})_{ijkl} &= 
 (\overline{\ell}_i \gamma_{\mu} \ell_j)(\overline{q}_k \gamma^{\mu} q_l),
 \\
 (\mathcal{O}^{(3)}_{\ell q})_{ijkl} &= 
 (\overline{\ell}_i \gamma_{\mu}\tau^I \ell_j)(\overline{q}_k \gamma^{\mu}\tau^I q_l),
 \\
 (\mathcal{O}_{qe})_{ijkl} &= 
 (\overline{q}_i\gamma_{\mu} q_j)(\overline{e}_{Rk} \gamma^{\mu} e_{Rl}),
 \\
 (\mathcal{O}_{\ell u})_{ijkl} &= 
 (\overline{\ell}_i \gamma_{\mu} \ell_j)(\overline{u}_{Rk} \gamma^{\mu} u_{Rl}),
 \\
 (\mathcal{O}_{eu})_{ijkl} &= 
 (\overline{e}_{Ri} \gamma_{\mu} e_{Rj})(\overline{u}_{Rk} \gamma^{\mu} u_{Rl}),
 \\
 (\mathcal{O}^{(1)}_{Hq})_{ij} &= 
 (H^\dagger i\overleftrightarrow{D_\mu} H) (\overline{q}_{i} \gamma^\mu q_{j}),
 \\
 (\mathcal{O}^{(3)}_{Hq})_{ij} &= 
 (H^\dagger i\overleftrightarrow{D_\mu^I} H) (\overline{q}_{i} \gamma^\mu \tau^I q_{j}),
 \\
 (\mathcal{O}_{Hu})_{ij} &= 
 (H^\dagger i\overleftrightarrow{D_\mu} H) (\overline{u}_{Ri} \gamma^\mu u_{Rj}),
\end{align}
where the derivative is
\begin{align}
    H^{\dagger}\overleftrightarrow{D}^I_{\mu}H =H^{\dagger}\tau^ID_{\mu}H -(D_{\mu}H)^{\dagger}\tau^I H,
\end{align}
and $\tau^I$ is the $SU(2)_L$ generator with flavor indices $i,j,k,l$. 
Since it is straightforward to extend the following analysis for the case of general $d_i\to d_j \ell\ell$ transitions, we do not specify the fermion flavors here and hereafter. 

The Higgs field gets a vacuum expectation value, $v = (\sqrt{2}G_F)^{-1/2} \simeq 246\GeV$. 
Below the EWSB scale, the LEFT Hamiltonian is defined as\footnote{Note that the normalization is the same as Ref.~\cite{Aebischer:2015fzz}, but is different from many of the literature such as Ref.~\cite{Ciuchini:2020gvn}}
\begin{align}
 \mathcal{H}_{\rm eff} &=
 -\frac{4G_F}{\sqrt{2}} \left[C_9 Q_9 +C_{10} Q_{10} \right],
 \label{eq:LEFT}
\end{align}
where the operators are
\begin{align}
 (Q_{9})_{ijkl} &= 
 \frac{e^2}{16\pi^2} (\overline{d}_i \gamma_\mu P_L d_j) (\overline{e}_k \gamma^\mu e_l), \\
 (Q_{10})_{ijkl} &= 
 \frac{e^2}{16\pi^2} (\overline{d}_i \gamma_\mu P_L d_j) (\overline{e}_k \gamma^\mu \gamma_5 e_l),
\end{align}
where the fields are regarded as Dirac fermions in the mass eigenstate basis. 
Let us also define the following operators to use in the following sections,
\begin{align}
 (\mathcal{Q}_{L})_{ijkl} &= (\overline{d}_i \gamma_\mu P_L d_j) (\overline{e}_k \gamma^\mu P_L e_l), \\
 (\mathcal{Q}_{R})_{ijkl} &= (\overline{d}_i \gamma_\mu P_L d_j) (\overline{e}_k \gamma^\mu P_R e_l), \\
 (\mathcal{Q}_{Z})_{ijkl} &= (-1+2s_W^2) (\mathcal{Q}_{L})_{ijkl} + 2s_W^2 (\mathcal{Q}_{R})_{ijkl},
\end{align}
where $s_W$ is the sine of the Weinberg angle. 
At the tree level, the matching conditions between the SMEFT and LEFT operators are obtained as (see, {\it e.g.}, Ref.~\cite{Aebischer:2015fzz})
\begin{align}
 (C_{9})_{ijkl}^{\rm tree} &= 
 \frac{\pi v^2}{\alpha} 
 \left\{
 (C_{qe})_{ijkl} + (C^{(1)}_{\ell q})_{klij} + (C^{(3)}_{\ell q})_{klij} 
 + \left[ (C^{(1)}_{Hq})_{ij} + (C^{(3)}_{Hq})_{ij} \right] \delta_{kl} (-1+4s_W^2)
 \right\},
 \label{eq:tree1} \\
 (C_{10})_{ijkl}^{\rm tree} &= 
 \frac{\pi v^2}{\alpha} 
 \left\{ (C_{qe})_{ijkl} - (C^{(1)}_{\ell q})_{klij} - (C^{(3)}_{\ell q})_{klij} 
 + \left[ (C^{(1)}_{Hq})_{ij} + (C^{(3)}_{Hq})_{ij} \right] \delta_{kl} \right\},
 \label{eq:tree2}
\end{align}
where $\alpha$ is the fine structure constant.
Note that the Wilson coefficients in both sides are evaluated at the EWSB scale.

\section{EW one-loop corrections}
\label{sec:EWoneloop}

\begin{figure}[thp]
\begin{center}
\begin{tabular}{ccc}
\subfigure[]{
\raisebox{0.6cm}{
\includegraphics[width=0.25\textwidth, bb= 0 0 118 49]{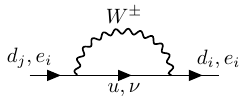}
}
} &
\subfigure[]{
\includegraphics[width=0.25\textwidth, bb= 0 0 127 94]{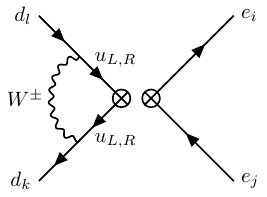}
} &
\subfigure[]{
\includegraphics[width=0.25\textwidth, bb= 0 0 141 94]{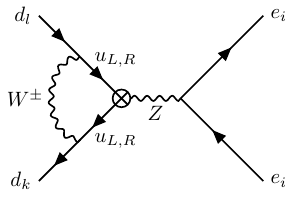}
}
\\
\subfigure[]{
\includegraphics[width=0.25\textwidth, bb= 0 0 129 94]{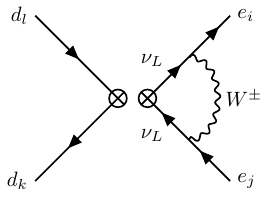}
} &
\subfigure[]{
\includegraphics[width=0.25\textwidth, bb= 0 0 126 100]{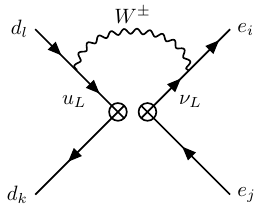}
} &
\subfigure[]{
\raisebox{0.3cm}{
\includegraphics[width=0.25\textwidth, bb= 0 0 111 62]{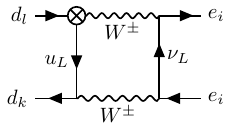}
}
}
\\
\subfigure[]{
\includegraphics[width=0.25\textwidth, bb= 0 0 143 94]{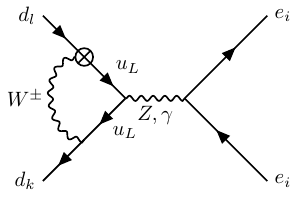}
} &
\subfigure[]{
\includegraphics[width=0.25\textwidth, bb= 0 0 139 94]{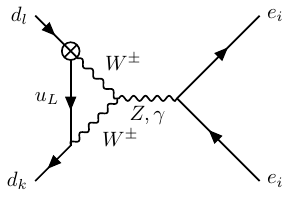}
} &
\subfigure[]{
\includegraphics[width=0.25\textwidth, bb= 0 0 139 94]{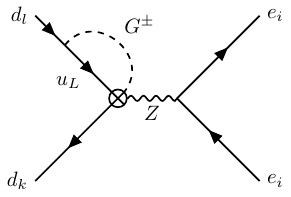}
}
\end{tabular}
\caption{
Feynman diagrams for the EW one-loop corrections.
Additionally, there are diagrams where the Nambu–Goldstone (NG) boson $G^\pm$, instead of $W^\pm$, is exchanged.
The vertices with a cross in a circle indicate SMEFT operators. 
In (a), the vertices are provided either by the SMEFT or the SM. 
For (f--h), the SMEFT vertex can also be assigned to the lower quark line.
Similarly, the NG boson can be attached to the lower quark instead of the upper one in (i).
}
\label{fig:diagrams}
\end{center}
\end{figure}

In this section, we calculate building blocks of the EW one-loop matching formulae, particularly paying attention to the contributions which change the quark flavors by the CKM matrix.
There are five types of $W$-boson loop diagrams;
i) self-energy corrections [Fig.~\ref{fig:diagrams}(a)],
ii) vertex corrections to right-handed up-type quarks [Fig.~\ref{fig:diagrams}(b,c)], 
iii) those to left-handed quarks/leptons [Fig.~\ref{fig:diagrams}(b\text{--}e)],
iv) box contributions [Fig.~\ref{fig:diagrams}(f)], 
and v) penguin contributions mediated by photon or $Z$ boson [Fig.~\ref{fig:diagrams}(g\text{--}i)].
We adopt the dimensional regularization scheme in the $R_{\xi}$ gauge by means of the {\tt FeynArts} \cite{Hahn:2000kx} and {\tt FormCalc} \cite{Hahn:1998yk} packages, where the SMEFT operators are implemented via {\tt Feynrules} \cite{Alloul:2013bka}.\footnote{
 In the analysis of four-Fermi operators, spurious fields are introduced to split them into renormalizable interactions to avoid a fermion chain problem (see the manual of {\tt FeynArts}).
}
Besides, we apply the on-shell renormalization condition.\footnote{
 The tadpole contributions are chosen to be zero.
}
In the calculations, all the quark masses are kept non-zero especially for dealing with the quark-flavor mixings.
Then, the external quark masses as well as those of leptons are approximated to be zero afterwards, because they are much smaller than the NP scale. 

\subsection{Self-energy corrections}

The wave-function renormalization must be taken into account to study the quark-flavor transitions as well as the gauge independence. 
In general, the inverse fermion propagator is expressed as 
\begin{align}
 S_{ij}^{-1} &= 
 (\slashed{p} - m_i) \delta_{ij} - \Sigma_{ij}(p).
\end{align}
Here, $\Sigma_{ij}$ is the unrenormalized self-energy amplitude, which is represented as
\begin{align}
\Sigma_{ij}(p) &= 
   \slashed{p}[P_L \Sigma_{ij}^L(p^2) + P_R \Sigma_{ij}^R(p^2)]
 + P_L \Sigma_{ij}^{DL}(p^2) + P_R \Sigma_{ij}^{DR}(p^2).
\end{align}
We denote a Dirac fermion in the mass eigenstates basis by $f_i$ with a flavor index, $i=1,2,3$.
The unrenormalized fields are written in terms of the renormalized ones as 
\begin{align}
 f_{L,R,i}^{(0)} 
 &= (Z_{ij}^{L,R})^{1/2} f_{L,R,j} 
 = \left(\delta_{ij} + \frac{1}{2} \delta Z_{ij}^{L,R}\right)f_{L,R,j},
 \\
 \bar{f}_{L,R,i}^{(0)} 
 &= (\overline{Z}_{ij}^{L,R})^{1/2} \bar{f}_{L,R,j}
 = \left(\delta_{ij} + \frac{1}{2} \delta \overline{Z}_{ij}^{L,R}\right)\bar{f}_{L,R,j}.
\end{align}

To determine the renormalization constants, we impose the on-shell conditions and follow the prescription which has been developed for a chiral theory in the presence of fermion mixings\cite{Aoki:1982ed,Bohm:1986rj,Denner:1990yz,Kniehl:1996bd,Pilaftsis:2002nc,Denner:1991kt}.
In particular, those for $i = j$ are obtained as
\begin{align}
 \delta Z_{ii}^{L} &= \Sigma_{ii}^L(m_i^2) + A + D + \alpha_i, \label{eq:ZLii} \\
 \delta Z_{ii}^{R} &= \Sigma_{ii}^R(m_i^2) - A + D + \alpha_i, \\
 \delta \overline{Z}_{ii}^{L} &= \Sigma_{ii}^L(m_i^2) - A + D - \alpha_i, \\
 \delta \overline{Z}_{ii}^{R} &= \Sigma_{ii}^R(m_i^2) + A + D - \alpha_i, \label{eq:ZRbii}
\end{align}
with 
\begin{align}
 A &= \frac{1}{2m_i} \left[ \Sigma_{ii}^{DR}(m_i^2) - \Sigma_{ii}^{DL}(m_i^2) \right], \\
 D &= 
   m_i^2 \left[ \Sigma_{ii}^{L\prime}(m_i^2)  + \Sigma_{ii}^{R\prime}(m_i^2)  \right]
 + m_i   \left[ \Sigma_{ii}^{DL\prime}(m_i^2) + \Sigma_{ii}^{DR\prime}(m_i^2) \right],
\end{align}
where $m_i$ is a mass of the external fermion $i$, and $\Sigma_{ii}^{\prime}(m_i^2) = \partial \Sigma_{ii}(p^2)/\partial p^2|_{p^2=m_i^2}$. 
Also, $\alpha_i$ is a free parameter, which cannot be fixed by the on-shell condition and does not affect physical observables, satisfying
\begin{align}
 \delta Z_{ii}^{L} + \delta Z_{ii}^{R} = \delta \overline{Z}_{ii}^{L} + \delta \overline{Z}_{ii}^{R} + 4 \alpha_i.
\end{align}
We choose $\alpha_i = 0$ in the following analysis. 
On the other hand, the flavor off-diagonal components $(i \neq j)$ are obtained as
\begin{align}
 \delta Z_{ij}^{L} &=
 \frac{2}{m_j^2 - m_i^2}
 \left[
   m_j^2 \Sigma_{ij}^L(m_j^2) 
 + m_i m_j \Sigma_{ij}^R(m_j^2) 
 + m_i \Sigma_{ij}^{DL}(m_j^2) 
 + m_j \Sigma_{ij}^{DR}(m_j^2) 
 \right], 
 \label{eq:ZLij} \\
 \delta Z_{ij}^{R} &=
 \frac{2}{m_j^2 - m_i^2}
 \left[
   m_j^2 \Sigma_{ij}^R(m_j^2) 
 + m_i m_j \Sigma_{ij}^L(m_j^2) 
 + m_i \Sigma_{ij}^{DR}(m_j^2) 
 + m_j \Sigma_{ij}^{DL}(m_j^2) 
 \right], \\
 \delta \overline{Z}_{ij}^{L} &=
 \frac{2}{m_i^2 - m_j^2}
 \left[
   m_i^2 \Sigma_{ij}^L(m_i^2) 
 + m_i m_j \Sigma_{ij}^R(m_i^2)  
 + m_i \Sigma_{ij}^{DL}(m_i^2) 
 + m_j \Sigma_{ij}^{DR}(m_i^2) 
 \right], \\
 \delta \overline{Z}_{ij}^{R} &=
 \frac{2}{m_i^2 - m_j^2}
 \left[
   m_i^2 \Sigma_{ij}^R(m_i^2) 
 + m_i m_j \Sigma_{ij}^L(m_i^2) 
 + m_i \Sigma_{ij}^{DR}(m_i^2) 
 + m_j \Sigma_{ij}^{DL}(m_i^2) 
 \right].
 \label{eq:ZRbij}
\end{align}

Let us show the results of the above renormalization constants for the down-type quarks explicitly.
The EW one-loop diagram is given in Fig.~\ref{fig:diagrams}(a). 
During the calculation, all the fermion masses expect for those of the neutrinos are kept non-zero particularly to deal with the quark-flavor mixings. 
We expand the results by a ratio of the external quark masses, $m_i/m_j$ for $i < j$, and focus on the leading contribution. 
Then, the external quark masses as well as the lepton ones are approximated to zero. 
When both of the $W$-boson vertices in the diagram are provided by the SM, the result becomes
\begin{align}
 \frac{1}{2} \delta Z_{ij}^{L}
 &=
 \frac{\alpha \lambda_{m'}^{ij}}{\pi s_W^2} 
 \left[ K_{ij}(x_{m'},\mu) - \Xi_1(x_{m'}) - \frac{1}{2} \Xi_0(\mu) \right],
 \label{eq:zdL}
 \\
 \frac{1}{2} \delta \overline{Z}_{ij}^{L}  
 &=
 \frac{\alpha \lambda_{m'}^{ij}}{\pi s_W^2} 
 \left[ \overline{K}_{ij}(x_{m'},\mu) - \Xi_1(x_{m'}) - \frac{1}{2} \Xi_0(\mu) \right],
 \label{eq:zdbL}
 \\
 \frac{1}{2} \delta Z_{ij}^{R} &= 0,~~~
 \frac{1}{2} \delta \overline{Z}_{ij}^{R} = 0,
 \label{eq:zdR}
\end{align}
where the results are summed over the flavor index $m'$.
Here, $\lambda^{ij}_{m'}  = V^{\ast}_{m'i}V_{m'j}$, $x_i = m_{ui}^2/M_W^2$, and $\mu$ is the renormalization scale.
Note that an identity $\sum_{m'} \lambda^{ij}_{m'} \, f = \delta_{ij} f$ is satisfied due to the CKM unitarity as long as $f$ is independent of $x_m$.
From Eqs.~\eqref{eq:ZLii}--\eqref{eq:ZRbii} and \eqref{eq:ZLij}--\eqref{eq:ZRbij}, the loop functions depend on the flavor indices $i,j$ as
\begin{align}
 K_{ij}(x,\mu) =
 \begin{cases}
 K_0(x,\mu),
 & (i = j) \\
 K_1(x,\mu),
 & (i < j) \\
 K_2(x,\mu),
 & (i > j) 
 \end{cases}
 ~~~~~
 \overline{K}_{ij}(x,\mu) =
 \begin{cases}
 K_0(x,\mu),
 & (i = j) \\
 K_2(x,\mu),
 & (i < j) \\
 K_1(x,\mu),
 & (i > j) 
 \end{cases}
 \label{eq:wtnK}
\end{align}
with
\begin{align}
 K_0(x,\mu) &= 
 - \frac{x}{32} \left[
 \ln\frac{\mu^2}{M_W^2}
 + \frac{3(x+1)}{2(x-1)} 
 - \frac{x^2-2x+4}{(x-1)^2}\ln x
 \right],
 \\
 K_1(x,\mu) &= 
 \frac{x}{16} \left[
 \ln\frac{\mu^2}{M_W^2} 
 + \frac{x-7}{2(x-1)} 
 - \frac{x^2-2x-2}{(x-1)^2} \ln x
 \right],
 \\
 K_2(x,\mu) &= 
 - \frac{x}{8} \left[
 \ln\frac{\mu^2}{M_W^2} 
 + 1 - \ln x
 \right].
\end{align}
It is noticed that they satisfy an identity,
\begin{align}
 K_1(x,\mu) + K_2(x,\mu) = 2 K_0(x,\mu).
 \label{eq:IdentityK}
\end{align}
On the other hand, the functions involving the gauge parameter $\xi_W$ are given by
\begin{align}
 \Xi_0(\mu) &= 
 - \frac{3}{16} + \frac{\xi_W}{8} \left[ \ln\frac{\mu^2}{M_W^2} + 1 - \ln\xi_W \right],
 \label{eq:Xi0}
 \\
 \Xi_1(x) &= 
 - \frac{x}{8} \xi_W \frac{\ln x - \ln \xi_W}{x-\xi_W}.
 \label{eq:Xi1}
\end{align}
Here, the terms independent of $x$ as well as those with $\xi_W$ are assembled in $\Xi_0(\mu)$.

In order to show the gauge-parameter cancellation in the next section, we also need the renormalization constants for the charged leptons. 
From Fig.~\ref{fig:diagrams}(a) involving the neutrino instead of the up-type quark, they become
\begin{align}
 \frac{1}{2} \delta Z_{ii}^{eL} &= 
 \frac{1}{2} \delta \overline{Z}_{ii}^{eL} = 
 \frac{\alpha}{\pi s_W^2} 
 \left[ -\frac{1}{2} \Xi_0(\mu) \right],
 \label{eq:zeL}
 \\
 \frac{1}{2} \delta Z_{ii}^{eR} &= 0,~~~
 \frac{1}{2} \delta \overline{Z}_{ii}^{eR} = 0,
\end{align}
which are obtained from Eqs.~\eqref{eq:zdL}--\eqref{eq:zdR} with $x_m \to 0$ and $V \to 1$.

It is pedagogical to show explicitly how the effective operators are corrected by the wave-function renormalizations. 
Suppose an effective operator is given by
\begin{align}
 \mathcal{L}_{\rm eff} &= 
 C_{ijkl} (\overline{e}_i \gamma_{\mu} P_I e_j)(\overline{d}_k \gamma^{\mu} P_J d_l),
\end{align}
the renormalization constants contribute as
\begin{align}
 \delta\mathcal{L}_{\rm eff} &= 
 \frac{1}{2} \delta \overline{Z}^{eI}_{i'i}
 C_{ijkl}
 (\overline{e}_{i'} \gamma_{\mu} P_I e_j)(\overline{d}_k \gamma^{\mu} P_J d_l)
 + \frac{1}{2} \delta Z^{eI}_{jj'}
 C_{ijkl}
 (\overline{e}_i \gamma_{\mu} P_I e_{j'})(\overline{d}_k \gamma^{\mu} P_J d_l)
 \notag \\ &~ 
 +
 \frac{1}{2} \delta \overline{Z}^J_{k'k}
 C_{ijkl}
 (\overline{e}_i \gamma_{\mu} P_I e_j)(\overline{d}_{k'} \gamma^{\mu} P_J d_l)
 + \frac{1}{2} \delta Z^J_{ll'}
 C_{ijkl}
 (\overline{e}_i \gamma_{\mu} P_I e_j)(\overline{d}_k \gamma^{\mu} P_J d_{l'}),
\end{align}
where the summation over the primed indices in each term is implicit. 
From Eq.~\eqref{eq:wtnK}, it is noticed that the quark renormalization constants differ by the flavor structure.
Therefore, we will derive one-loop matching formulae in each case.\footnote{
 We have focused on the SMEFT operators in the mass eigenstate basis. 
 If they are defined in the {\it gauge} eigenstate basis, the matching conditions onto the LEFT operators in the {\it mass} eigenstate basis should involve additional EW corrections to change the basis (see, {\it e.g.}, Ref.~\cite{Kniehl:1996bd}).
}

In the above results, all the $W$-boson interactions are set by the SM. 
Additionally, they are provided by the operator $\mathcal{O}^{(3)}_{Hq}$ as
\begin{align}
 \mathcal{L}_{\rm eff} &= 
 \frac{gv^2}{\sqrt{2}} 
 (C^{(3)}_{Hq})_{kl}
 \Big[
  V_{m'k}   \overline{u}_{m'} \gamma^\mu P_L d_{l}  W_\mu + 
  V_{m'l}^* \overline{d}_k    \gamma^\mu P_L u_{m'} W_\mu^\dagger 
 \Big],
 \label{eq:WHQ3}
\end{align}
where $g$ is the $SU(2)_L$ gauge coupling. 
There are also NG-boson interactions.\footnote{
  The complete Feynman rules in the $R_{\xi}$ gauge are found, {\it e.g.}, in Ref.~\cite{Dedes:2017zog}. 
}
By replacing one of the vertices with it in Fig.~\ref{fig:diagrams}(a), the renormalization constants for $i \neq j$ become
\begin{align}
 \frac{1}{2} \delta Z_{ij}^{L\,(HQ3)}
 &=
 \frac{\alpha v^2}{\pi s_W^2} 
 \left[ \lambda_{m'}^{j'j} (C^{(3)}_{Hq})_{ij'} + \lambda_{m'}^{ii'} (C^{(3)}_{Hq})_{i'j} \right]
 \big[ K_1(x_{m'},\mu) - \Xi_1(x_{m'}) \big],~~~(i < j)
 \\
 \frac{1}{2} \delta \overline{Z}_{ij}^{L\,(HQ3)}  
 &=
 \frac{\alpha v^2}{\pi s_W^2} 
 \left[ \lambda_{m'}^{j'j} (C^{(3)}_{Hq})_{ij'} + \lambda_{m'}^{ii'} (C^{(3)}_{Hq})_{i'j} \right]
 \big[ K_1(x_{m'},\mu) - \Xi_1(x_{m'}) \big],~~~(i > j)
\end{align}
and the others are zero.
Here, the summation is made over the primed indices. 
They contribute to $d_k \to d_l \ell\ell$ $(k \neq l)$ by combined with the SM amplitude, $d_{m} \to d_{m} \ell\ell$, which is generated by exchanging the $Z$ boson.
Consequently, the effective operator becomes
\begin{align}
 \delta\mathcal{L}_{\rm eff} &= 
 \frac{\alpha}{\pi s_W^2}
 \lambda_{m'}^{k'k} 
 \left[ 1 - \frac{2}{3}s_W^2 \right]
 \left[ K_1(x_{m'},\mu) - \Xi_1(x_{m'}) \right] 
 (C^{(3)}_{Hq})_{kl} (\mathcal{Q}_{Z})_{k'lii}
 \notag \\ &~~
 +
 \frac{\alpha}{\pi s_W^2}
 \lambda_{n'}^{ll'} 
 \left[ 1 - \frac{2}{3}s_W^2 \right]
 \left[ K_1(x_{n'},\mu) - \Xi_1(x_{n'}) \right] 
 (C^{(3)}_{Hq})_{kl} (\mathcal{Q}_{Z})_{kl'ii}.
 \label{eq:wtnKHQ3}
\end{align}

\subsection{Vertex corrections to right-handed up-type quarks}

In this and following subsections, we consider radiative corrections to vertices.
Let us first consider an effective operator,
\begin{align}
 \mathcal{L}_{\rm eff} &= (\overline{u}_m \gamma_{\mu} P_R u_n) \mathcal{Q}^\mu_{mn},
\end{align}
where the right-handed up-type quarks are involved, and $\mathcal{Q}^\mu_{mn}$ is a remnant of the operator such as
$\mathcal{Q}^\mu_{mn} = C_{ijmn} (\overline{e}_i \gamma^{\mu} P_{L,R} e_j)$ with the Wilson coefficient $C_{ijmn}$.
Note that the operator is shown in the mass eigenstate basis.
For the vertex correction to the quarks as shown in Fig.~\ref{fig:diagrams}(b,c), we obtain
\begin{align}
 \delta\mathcal{L}_{\rm eff} &= 
 \frac{\alpha}{\pi s_W^2}
 V^*_{m'k'} V_{n'l'} \,
 I_1(x_{m'},x_{n'},\mu) \,
 (\overline{d}_{k'} \gamma_{\mu} P_L d_{l'}) \mathcal{Q}^\mu_{m'n'}.
 \label{eq:ve1}
\end{align}
The result is summed over the primed indices.
Here, the loop function is
\begin{align}
 I_1(x,y,\mu) &=
 \frac{\sqrt{xy}}{16} \left[ 
   \ln\frac{\mu^2}{M_W^2}
 + \frac{1}{2}
 - \frac{x(x-4)}{(x-1)(x-y)} \ln x 
 + \frac{y(y-4)}{(y-1)(x-y)} \ln y
 \right],
\end{align}
which is symmetric for exchanging $x$ and $y$.
We also obtain
\begin{align}
 I_1(x,\mu) &\equiv I_1(x,x,\mu) = 
 \frac{x}{16} \left[ 
   \ln\frac{\mu^2}{M_W^2}
 - \frac{x-7}{2(x-1)} 
 - \frac{x^2-2x+4}{(x-1)^2} \ln x 
 \right].
\end{align}
Although the calculation was performed in the $R_{\xi}$ gauge, the result is independent of the gauge parameter.
It should be clarified again that the result is represented by the fields in the mass eigenstate basis.

\subsection{Vertex corrections to left-handed quarks/leptons}

In the down basis, the left-handed quarks in an effective operator are expanded as
\begin{align}
 \mathcal{L}_{\rm eff} &= (\overline{q}_k \gamma_{\mu} P_L q_l) \mathcal{Q}^\mu_{kl}
 = V_{m'k}V_{n'l}^{\ast}(\overline{u}_{m'} \gamma_{\mu} P_L u_{n'}) \mathcal{Q}^\mu_{kl} + \cdots,
 \label{eq:qexpansion}
\end{align}
where the terms with the down-type quarks are omitted, and the summation over $m',n'$ is implicit. 
From Fig.~\ref{fig:diagrams}(b,c), the radiative correction to the up-type quarks yields
\begin{align}
 \delta\mathcal{L}_{\rm eff} &= 
 \frac{\alpha}{\pi s_W^2} \lambda^{k'k}_{m'} \lambda^{ll'}_{n'}
 \big[ J_2(x_{m'},x_{n'}) + \Xi_1(x_{m'}) + \Xi_1(x_{n'}) + \Xi_0(\mu) \big]
 (\overline{d}_{k'} \gamma_{\mu} P_L d_{l'}) \mathcal{Q}^\mu_{kl},
 \label{eq:ve2}
\end{align}
where the result is summed over $m',n',k',l'$. 
The loop function is
\begin{align}
 J_2(x,y) &= \frac{xy}{8} \frac{\ln x - \ln y}{x-y},
\end{align}
which is symmetric for $x \leftrightarrow y$.
Also, one obtains $J_2(x) \equiv J_2(x,x) = x/8$.
It is noticed that there are four CKM matrix elements in Eq.~\eqref{eq:ve2};
two of them come from Eq.~\eqref{eq:qexpansion}, and the others from the $W$-boson interactions. 
The result depends on the gauge parameter via $\Xi_{0,1}(x)$, whose cancellation will be shown in the next section.

For the effective operator involving the neutrinos, 
\begin{align}
 \mathcal{L}_{\rm eff} &= (\overline{\nu}_i \gamma_{\mu}P_L \nu_j) \mathcal{Q}^\mu_{ij},
\end{align}
the $W$-boson loop diagram shown in Fig.~\ref{fig:diagrams}(d) gives
\begin{align}
 \delta\mathcal{L}_{\rm eff} &= 
 \frac{\alpha}{\pi s_W^2} \,
 \Xi_0(\mu) \,
 (\overline{e}_i \gamma_{\mu} P_L e_j) \mathcal{Q}^\mu_{ij},
 \label{eq:ve3}
\end{align}
which is necessary to see the gauge independence in the next section. 

For the effective operator $\mathcal{O}^{(3)}_{\ell q}$, one also needs to take account of radiative corrections to the following type of the operator,
\begin{align}
 \mathcal{L}_{\rm eff} &= 
 C_{ijkl} V_{m'k} (\overline{e}_i \gamma_{\mu} P_L \nu_j)(\overline{u}_{m'} \gamma^{\mu} P_L d_l),
\end{align}
where the summation over $m'$ is implicit. 
By exchanging the $W$ boson between the neutrino and the up-type quark as shown in Fig.~\ref{fig:diagrams}(e), one obtains
\begin{align}
 \delta\mathcal{L}_{\rm eff} &= 
 \frac{\alpha}{\pi s_W^2}\lambda^{k'k}_{m'} \,
 \big[ \Xi_1(x_{m'}) + \Xi_0(\mu) \big] \,
 C_{ijkl} \,
 (\overline{e}_i \gamma_{\mu} P_L e_j)(\overline{d}_{k'} \gamma^{\mu} P_L d_l).\label{eq:ve4}
\end{align}
Here, the result is summed over the indices, $m',k'$.

\subsection{Other box and penguin contributions}
\label{sec:otherBP}

The operator $\mathcal{O}^{(3)}_{Hq}$ contributes to the semileptonic decays by exchanging the SM bosons between the operator and leptons. 
A part of the $Z$-penguin contributions are evaluated by the above self-energy/vertex corrections. 
Besides, there are contributions from the box, photon-penguin, and additional $Z$-penguin diagrams as shown in Fig.~\ref{fig:diagrams}(f--i).
In this article, we focus on the EW one-loop contributions which change their quark flavors by the CKM matrix. 
Hence, we need to consider only these diagrams.\footnote{
 Similarly, the above self-energy/vertex corrections are sufficient for the purpose of studying the flavor-changing EW corrections for $\mathcal{O}^{(1)}_{Hq}$.
 In Appendix \ref{sec:nonfl}, we show the gauge invariance of the $W$-boson loop contributions  which do not change the quark flavors by taking additional diagrams into account.
}

The box contribution is generated through the $W$ (NG) boson interaction in $\mathcal{O}^{(3)}_{Hq}$. From Fig.~\ref{fig:diagrams}(f), it is estimated as
\begin{align}
 \delta\mathcal{L}_{\rm eff} &= 
 - \frac{\alpha}{\pi s_W^2}
 \lambda_{m'}^{k'k} 
 \big[ B(x_{m'}) + \Xi_2(x_{m'}) \big] 
 (C^{(3)}_{Hq})_{kl} 
 (\mathcal{Q}_{L})_{k'lii}
 \notag \\ &~~~
 - \frac{\alpha}{\pi s_W^2}
 \lambda_{n'}^{ll'} 
 \big[ B(x_{n'}) + \Xi_2(x_{n'}) \big] 
 (C^{(3)}_{Hq})_{kl} 
 (\mathcal{Q}_{L})_{kl'ii},
 \label{eq:box}
\end{align}
where the result is summed over $m',k'$ ($n',l'$) in the first (second) line, and the terms without the CKM flavor transitions are ignored here and hereafter in this subsection.
The loop functions are defined as
\begin{align}
 B(x) &= 
 \frac{3}{16}x \left[ 
   \frac{1}{x-1} 
 - \frac{1}{(x-1)^2} \ln x 
 \right],
 \\
 \Xi_2(x) &= 
 - \frac{3}{8} x \frac{\xi_W}{x-\xi_W} \left[ \frac{1}{6} - \frac{\ln x}{x-1} + \frac{\ln \xi_W}{\xi_W-1} \right]
 - \frac{\xi_W}{2(x-\xi_W)} \Xi_1(x).
\end{align}

Similarly, the photon penguin diagrams shown in Fig.~\ref{fig:diagrams}(g,h) are estimated as
\begin{align}
 \delta\mathcal{L}_{\rm eff} &= 
 - \frac{\alpha}{\pi s_W^2}
 \lambda_{m'}^{k'k} 
 \big[ s_W^2 D(x_{m'}) - 2 s_W^2 \Xi_2(x_{m'}) \big] 
 (C^{(3)}_{Hq})_{kl} 
 \big[ (\mathcal{Q}_{L})_{k'lii} + (\mathcal{Q}_{R})_{k'lii} \big]
 \notag \\ &~~~
 - \frac{\alpha}{\pi s_W^2}
 \lambda_{n'}^{ll'} 
 \big[ s_W^2 D(x_{n'}) - 2 s_W^2 \Xi_2(x_{n'}) \big] 
 (C^{(3)}_{Hq})_{kl} 
 \big[ (\mathcal{Q}_{L})_{kl'ii} + (\mathcal{Q}_{R})_{kl'ii} \big],
 \label{eq:phpen}
\end{align}
where the result is summed over $m',k'$ ($n',l'$) in the first (second) line.
The loop function is defined as
\begin{align}
 D(x) = 
 -\frac{2}{9} \ln x
 -\frac{x}{72} \left[ 
   \frac{82x^2-151x+63}{(x-1)^3} 
 - \frac{10x^3+59x^2-138x+63}{(x-1)^4} \ln x 
 \right].
\end{align}

Finally, the $Z$-penguin contributions are generated from Fig.~\ref{fig:diagrams}(g--i), where the $Z$ boson instead of the photon is exchanged.
The result is
\begin{align}
 \delta\mathcal{L}_{\rm eff} &= 
 - \frac{\alpha}{\pi s_W^2}
 \lambda_{m'}^{k'k} 
 \bigg[
   J_3(x_{m'},\mu) + \Xi_2(x_{m'}) - 3 \Xi_1(x_{m'})
 \notag \\ &\qquad\qquad\qquad
   - \frac{2}{3} s_W^2 
   \big[ K_1(x_{m'},\mu) - \Xi_1(x_{m'}) \big]
 \bigg] 
 (C^{(3)}_{Hq})_{kl} \mathcal{Q}_{k'lii},
 \notag \\ &~~~
 - \frac{\alpha}{\pi s_W^2}
 \lambda_{n'}^{ll'} 
 \bigg[
   J_3(x_{n'},\mu) + \Xi_2(x_{n'}) - 3 \Xi_1(x_{n'})
 \notag \\ &\qquad\qquad\qquad
   - \frac{2}{3} s_W^2 
   \big[ K_1(x_{n'},\mu) - \Xi_1(x_{n'}) \big]
 \bigg] 
 (C^{(3)}_{Hq})_{kl} \mathcal{Q}_{kl'ii},
 \label{eq:zphpen}
\end{align}
where the summation over the primed indices is implicit. 
The loop function is defined as
\begin{align}
     J_3(x,\mu) = 
 -\frac{3}{16}x \left[ 
   \ln\frac{\mu^2}{M_W^2}
 + \frac{x+3}{2(x-1)} 
 - \frac{x^2+1}{(x-1)^2} \ln x 
 \right].
\end{align}

\section{Result}

Below the EWSB scale, the EW bosons $(W, Z, H)$ and the top quark $(t)$ are decoupled from the effective theory.
In this section, we provide the matching formulae between the SMEFT operators and the LEFT ones for $d_i\to d_j \ell\ell$ at the EW one-loop level.\footnote{
 In this article, we ignore QCD long-distance effects on light-quark loops, which will be studied in future.
} 
We particularly pay attention to the contributions in which the quark flavors are changed due to the CKM matrix. 
The renormalization scale $\mu$ in the previous section is set to be the EWSB scale.

\subsection{$(C_{e u})_{ijkl}$, $(C_{\ell u})_{ijkl}$}

Let us first consider the four-Fermi interactions. 
The Wilson coefficient $(C_{e u})_{ijkl}$ is given by the effective Lagrangian,
\begin{align}
 \mathcal{L}_{\rm eff} &= 
 (C_{e u})_{ijkl}
 (\overline{e}_i \gamma_{\mu}P_R e_j)(\overline{u}_k \gamma^{\mu} P_R u_l),
\end{align}
which contributes to $d_i\to d_j \ell\ell$ through the vertex correction with respect to the right-handed up-type quarks.
From Eq.~\eqref{eq:ve1}, we obtain
\begin{align}
 \delta\mathcal{L}_{\rm eff} &= 
 \frac{\alpha}{\pi s_W^2}
 V^*_{m'k'} V_{n'l'} \,
 I_1(x_{m'},x_{n'},\mu) \,
 (C_{e u})_{ijm'n'}
 (\mathcal{Q}_{R})_{k'l'ij}.
\end{align}
There are no additional $W$-boson loop contributions.
Thus, the LEFT coefficients in Eq.~\eqref{eq:LEFT} are derived as
\begin{align}
 (C_9)^{\rm EW}_{ijkl} &= (C_{10})^{\rm EW}_{ijkl} = 
 \frac{v^2}{s_W^2} V^*_{m'i} V_{n'j} \, I_1(x_{m'},x_{n'},\mu) \, (C_{e u})_{klm'n'},\label{eq:euloop}
\end{align}
where the result is summed over $m',n'$.
Here and hereafter, both the SMEFT and LEFT operators are evaluated at the EWSB scale $\mu$.
Since the loop function $I_1(x,y,\mu)$ is proportional to $\sqrt{xy}$, the contribution at $m'=n'=3$, {\it i.e.}, the top-quark contribution, is likely to dominate as long as it is non-vanishing. 
This contribution is found to be consistent with the result in Ref.~\cite{Aebischer:2015fzz}.

Next, the effective operator of $(C_{\ell u})_{ijkl}$ is expanded as
\begin{align}
 \mathcal{L}_{\rm eff} &= 
 (C_{\ell u})_{ijkl}
 (\overline{\ell}_i \gamma_{\mu} \ell_j)(\overline{u}_{Rk} \gamma^{\mu} u_{Rl}) 
 \notag \\ &= 
 (C_{\ell u})_{ijkl}
 \left[ (\overline{\nu}_i \gamma_{\mu}P_L \nu_j) + (\overline{e}_i \gamma_{\mu}P_L e_j) \right]
 (\overline{u}_k \gamma^{\mu}P_R u_l).
\end{align}
From the vertex correction to the quarks in Eq.~\eqref{eq:ve1}, the EW corrections become
\begin{align}
 \delta\mathcal{L}_{\rm eff} &= 
 \frac{\alpha}{\pi s_W^2}
 V^*_{m'k'} V_{n'l'} \,
 I_1(x_{m'},x_{n'},\mu) \,
 (C_{\ell u})_{ijm'n'}
 (\mathcal{Q}_{L})_{k'l'ij}.
\end{align}
Although there are radiative corrections to the lepton vertex and self-energies, it is easily found from Eqs.~\eqref{eq:zeL} and \eqref{eq:ve3} that these contributions cancel in the sum.
Thus, the matching conditions are derived as
\begin{align}
 (C_9)^{\rm EW}_{ijkl} &= - (C_{10})^{\rm EW}_{ijkl} =
 \frac{v^2}{s_W^2}  V^*_{m'i} V_{n'j} \, I_1(x_{m'},x_{n'},\mu) \, (C_{\ell u})_{klm'n'},\label{eq:cluloop}
\end{align}
where the result is summed over $m',n'$.
If we pick up $m'=n'=3$, the results in Ref.~\cite{Aebischer:2015fzz} are reproduced.

\subsection{$(C_{q e})_{ijkl}$}

In the down basis, the left-handed quarks are expanded as
\begin{align}
 \mathcal{L}_{\rm eff} &= 
 (C_{q e})_{ijkl}
 (\overline{q}_i\gamma_{\mu} q_j)(\overline{e}_{Rk} \gamma^{\mu} e_{Rl})
 \notag \\ &= 
 (C_{q e})_{ijkl}
 \left[ 
 V_{m'i} V^*_{n'j} (\overline{u}_{m'} \gamma_{\mu} P_L u_{n'})
 + (\overline{d}_i \gamma_{\mu} P_L d_j)
 \right]
 (\overline{e}_k \gamma^{\mu}P_R e_l).
\end{align}
Equations \eqref{eq:zdL}, \eqref{eq:zdbL}, and \eqref{eq:ve2} contribute to the one-loop corrections as
\begin{align}
 \delta\mathcal{L}_{\rm eff} &= 
 \frac{\alpha}{\pi s_W^2}
 \lambda_{m'}^{i'i} \lambda_{n'}^{jj'} 
 \big[ J_2(x_{m'},x_{n'}) + \Xi_1(x_{m'}) + \Xi_1(x_{n'}) + \Xi_0(\mu) \big]
 (C_{q e})_{ijkl}
 (\mathcal{Q}_{R})_{i'j'kl}
 \notag \\ &~~~
 + \frac{1}{2} \delta \overline{Z}^L_{i'i} (C_{q e})_{ijkl} (\mathcal{Q}_{R})_{i'jkl}
 + \frac{1}{2} \delta Z^L_{jj'} (C_{q e})_{ijkl} (\mathcal{Q}_{R})_{ij'kl}
 \notag \\ &=
 \frac{\alpha}{\pi s_W^2} \lambda_{m'}^{i'i} \lambda_{n'}^{jj'} 
 \big[ J_2(x_{m'},x_{n'}) + \overline{K}_{i'i}(x_{m'},\mu) + K_{jj'}(x_{n'},\mu) \big]
 (C_{q e})_{ijkl}
 (\mathcal{Q}_{R})_{i'j'kl}.
\end{align}
In the second equality, we used $\sum_m \lambda^{ij}_m \, f = \delta_{ij} f$ for $f$ being independent of $x_m$.
It is found that all the gauge parameters via $\Xi_{0,1}$ are canceled out completely. 

By matching the result onto the LEFT, one obtains
\begin{align}
 (C_9)^{\rm EW}_{ijkl} &= (C_{10})^{\rm EW}_{ijkl} 
 \notag \\ &=
 \frac{v^2}{s_W^2} \lambda_{m'}^{ii'} \lambda_{n'}^{j'j} 
 \Big[ J_2(x_{m'},x_{n'}) + \overline{K}_{ii'}(x_{m'},\mu) + K_{j'j}(x_{n'},\mu) \Big]
 (C_{q e})_{i'j'kl}.
\end{align}
The result is summed over the primed indices, as understood from the previous section. 
As shown in Eq.~\eqref{eq:wtnK}, $K_{ij}(x,\mu)$ and $\overline{K}_{ij}(x,\mu)$ depend on the flavor structure.
The results are explicitly shown as
\begin{align}
 (C_9)^{\rm EW}_{ijkl} &= (C_{10})^{\rm EW}_{ijkl} 
 \notag \\ &=
 \frac{v^2}{s_W^2} \lambda_{m'}^{ii'} \lambda_{n'}^{j'j} \, (C_{q e})_{i'j'kl} \times
 \begin{cases}
 I_2(x_{m'},x_{n'},\mu),
 & (i' = i, j' = j) \\
 \left[ J_2(x_{m'},x_{n'}) + K_1(x_{m'},\mu) \right],
 & (i' < i, j' = j) \\
 \left[ J_2(x_{m'},x_{n'}) + K_2(x_{m'},\mu) \right],
 & (i' > i, j' = j) \\
 \left[ J_2(x_{m'},x_{n'}) + K_1(x_{n'},\mu) \right],
 & (i' = i, j' < j) \\
 \left[ J_2(x_{m'},x_{n'}) + K_2(x_{n'},\mu) \right],
 & (i' = i, j' > j) \\
 J_2(x_{m'},x_{n'}),
 & (i' \neq i, j' \neq j) 
 \end{cases}
 \label{eq:ResultCqe}
\end{align}
where the loop function is defined as
\begin{align}
 I_2(x,y,\mu) &= J_2(x,y) + K_0(x,\mu) + K_0(y,\mu),
\end{align}
which satisfies $I_2(x,x,\mu) = -I_1(x,\mu)$.

In order to compare the result with the literature, let us specify the operator to the quark-flavor universal case,
\begin{align}
 (C_{q e})_{ijkl} = (C_{q e})_{kl} \delta_{ij}.
\end{align}
Since an identity $\sum_{i',j'}\lambda_{m'}^{ii'} \lambda_{n'}^{j'j} \delta_{i'j'} = \lambda_{m'}^{ij} \delta_{m'n'}$ holds due to the CKM unitarity, the above result is simplified as
\begin{align}
 (C_9)^{\rm EW}_{ijkl} &= (C_{10})^{\rm EW}_{ijkl} 
 \notag \\ &=
 \frac{v^2}{s_W^2} \lambda_{m'}^{ij} 
 \Big[ J_2(x_{m'},x_{m'}) + \overline{K}_{ij}(x_{m'},\mu) + K_{ij}(x_{m'},\mu) \Big]
 (C_{q e})_{kl}
 \notag \\ &=
 -\frac{v^2}{s_W^2} \lambda_{m'}^{ij} 
 I_1(x_{m'},\mu)
 (C_{q e})_{kl},
 \label{eq:ResultCqeUniversal}
\end{align}
where the summation over $m'$ is implicit.
In the last equation, the identity in Eq.~\eqref{eq:IdentityK} was used to simplify the loop functions. 
Consequently, although the results of the loop function in the general case \eqref{eq:ResultCqe} depend on the flavor indices $i,j$, those in the flavor-universal case \eqref{eq:ResultCqeUniversal} become insensitive to them.\footnote{
 This mechanism is similar to the case of the EW corrections to the SM gauge interactions.
} 
Since $I_1(x,\mu)$ is proportional to $x$, it is dominated by the top-quark contribution, $m'=3$.
Then, the result becomes consistent with those in Ref.~\cite{Hurth:2019ula}, where the flavor-universal case was studied. 
Let us stress again that the result, however, differs from it when the SMEFT operator is not flavor-universal.

\subsection{$(C^{(1)}_{\ell q})_{ijkl}$}

The analysis on $(C^{(1)}_{\ell q})_{ijkl}$ is very similar to that of $(C_{q e})_{ijkl}$.
The left-handed quarks (and leptons) are expanded as
\begin{align}
 \mathcal{L}_{\rm eff} &= 
 (C^{(1)}_{\ell q})_{ijkl}
 (\overline{\ell}_i \gamma_{\mu} \ell_j)(\overline{q}_k \gamma^{\mu} q_l)
 \notag \\ &= 
 (C^{(1)}_{\ell q})_{ijkl}
 (\overline{e}_i \gamma_{\mu}P_L e_j)
 \left[ 
 V_{m'k} V^*_{n'l} (\overline{u}_{m'} \gamma^{\mu} P_L u_{n'})
 + (\overline{d}_k \gamma^{\mu} P_L d_l)
 \right]
 + \ldots,
\end{align}
where the terms including the neutrinos are omitted.
From Eqs.~\eqref{eq:zdL}, \eqref{eq:zdbL}, and \eqref{eq:ve2}, we obtain
\begin{align}
 \delta\mathcal{L}_{\rm eff} &= 
 \frac{\alpha}{\pi s_W^2}
 \lambda_{m'}^{k'k} \lambda_{n'}^{ll'} 
 \big[ J_2(x_{m'},x_{n'}) + \Xi_1(x_{m'}) + \Xi_1(x_{n'}) + \Xi_0(\mu) \big]
 (C^{(1)}_{\ell q})_{ijkl}
 (\mathcal{Q}_{L})_{k'l'ij}
 \notag \\ &~~~
 + \frac{1}{2} \delta \overline{Z}^L_{k'k} (C^{(1)}_{\ell q})_{ijkl} (\mathcal{Q}_{L})_{k'lij}
 + \frac{1}{2} \delta Z^L_{ll'} (C^{(1)}_{\ell q})_{ijkl} (\mathcal{Q}_{L})_{kl'ij}
 \notag \\ &=
 \frac{\alpha}{\pi s_W^2} \lambda_{m'}^{k'k} \lambda_{n'}^{ll'} 
 \big[ J_2(x_{m'},x_{n'}) + \overline{K}_{k'k}(x_{m'},\mu) + K_{ll'}(x_{n'},\mu) \big]
 (C^{(1)}_{\ell q})_{ijkl}
 (\mathcal{Q}_{L})_{k'l'ij}.
\end{align}
This is obviously gauge invariant.
Note that the radiative corrections to the leptons (neutrinos) are canceled similarly to the case of $(C_{\ell u})_{ijkl}$.
Thus, the matching formulae become
\begin{align}
 (C_9)^{\rm EW}_{ijkl} &= - (C_{10})^{\rm EW}_{ijkl} 
 \notag \\ &=
 \frac{v^2}{s_W^2} \lambda_{m'}^{ii'} \lambda_{n'}^{j'j} 
 \Big[ J_2(x_{m'},x_{n'}) + \overline{K}_{ii'}(x_{m'},\mu) + K_{j'j}(x_{n'},\mu) \Big]
 (C^{(1)}_{\ell q})_{kli'j'}
 \notag \\ &=
 \frac{v^2}{s_W^2} \lambda_{m'}^{ii'} \lambda_{n'}^{j'j} \, (C^{(1)}_{\ell q})_{kli'j'} \times
 \begin{cases}
 I_2(x_{m'},x_{n'},\mu),
 & (i' = i, j' = j) \\
 \left[ J_2(x_{m'},x_{n'}) + K_1(x_{m'},\mu) \right],
 & (i' < i, j' = j) \\
 \left[ J_2(x_{m'},x_{n'}) + K_2(x_{m'},\mu) \right],
 & (i' > i, j' = j) \\
 \left[ J_2(x_{m'},x_{n'}) + K_1(x_{n'},\mu) \right],
 & (i' = i, j' < j) \\
 \left[ J_2(x_{m'},x_{n'}) + K_2(x_{n'},\mu) \right],
 & (i' = i, j' > j) \\
 J_2(x_{m'},x_{n'}).
 & (i' \neq i, j' \neq j) 
 \end{cases}\label{eq:c1lqloop}
\end{align}

When we consider the quark-flavor universal case to compare the results with those in Ref.~\cite{Hurth:2019ula}, the operator is specified as
\begin{align}
 (C^{(1)}_{\ell q})_{ijkl} = (C^{(1)}_{\ell q})_{ij} \delta_{kl}.
\end{align}
Then, the matching conditions become
\begin{align}
 (C_9)^{\rm EW}_{ijkl} &= - (C_{10})^{\rm EW}_{ijkl} 
 \notag \\ &=
 \frac{v^2}{s_W^2} \lambda_{m'}^{ij} 
 \Big[ J_2(x_{m'},x_{m'}) + \overline{K}_{ij}(x_{m'},\mu) + K_{ij}(x_{m'},\mu) \Big]
 (C^{(1)}_{\ell q})_{kl}
 \notag \\ &=
 -\frac{v^2}{s_W^2} \lambda_{m'}^{ij} 
 I_1(x_{m'},\mu)
 (C^{(1)}_{\ell q})_{kl},
\end{align}
where the summation over $m'$ is implicit.
The result is dominated by the top-quark contribution, and becomes consistent with those in Ref.~\cite{Hurth:2019ula} by selecting $m'=3$.

\subsection{$(C^{(3)}_{\ell q})_{ijkl}$}

In order to evaluate the one-loop corrections associated with $(C^{(3)}_{\ell q})_{ijkl}$ and show the gauge invariance for this contribution, we expand the SMEFT Lagrangian as 
\begin{align}
 \mathcal{L}_{\rm eff} &= 
 (C^{(3)}_{\ell q})_{ijkl}
 (\overline{\ell}_i \gamma_{\mu} \tau^I \ell_j)(\overline{q}_k \gamma^{\mu} \tau^I q_l)
 \notag \\ &= 
 (C^{(3)}_{\ell q})_{ijkl}
 \Big\{ 
 \left[
   (\overline{\nu}_i \gamma_{\mu}P_L \nu_j)
 - (\overline{e}_i \gamma_{\mu} P_L e_j)
 \right]
 \left[
   V_{m'k} V^*_{n'l}
   (\overline{u}_{m'} \gamma^{\mu} P_L u_{n'})
 - (\overline{d}_k \gamma^{\mu} P_L d_l)
 \right]
 \notag \\ &\qquad\qquad\qquad
 + 2 V_{m'k} 
 (\overline{e}_i \gamma_{\mu}P_L \nu_j)
 (\overline{u}_{m'} \gamma^{\mu} P_L d_l)
 + 2 V^*_{n'l} 
 (\overline{\nu}_i \gamma_{\mu} P_L e_j)
 (\overline{d}_k \gamma^{\mu} P_L u_{n'})
 \Big\}.
\end{align}
The EW corrections to each term are provided by Eqs.~\eqref{eq:zdL}, \eqref{eq:zdbL}, \eqref{eq:zeL}, \eqref{eq:ve2}, \eqref{eq:ve3}, and \eqref{eq:ve4}.
As a result, we obtain
\begin{align}
 \delta\mathcal{L}_{\rm eff} &= 
 - \frac{\alpha}{\pi s_W^2}
 \lambda_{m'}^{k'k} \lambda_{n'}^{ll'}
 \Big[ J_2(x_{m'},x_{n'}) + \Xi_1(x_{m'}) + \Xi_1(x_{n'}) + \Xi_0(\mu) \Big]
 (C^{(3)}_{\ell q})_{ijkl}
 (\mathcal{Q}_{L})_{k'l'ij}
 \notag \\ &~~~
 - \frac{\alpha}{\pi s_W^2} \,
 \Xi_0(\mu) \,
 (C^{(3)}_{\ell q})_{ijkl}
 (\mathcal{Q}_{L})_{klij}
 \notag \\ &~~~
 +
 \frac{2\alpha}{\pi s_W^2}
 \Big[ \Xi_1(x_{m'}) + \Xi_0(\mu) \Big]
 (C^{(3)}_{\ell q})_{ijkl}
 \left[
  \lambda_{m'}^{k'k}
  (\mathcal{Q}_{L})_{k'lij}
  +
  \lambda_{m'}^{ll'}
  (\mathcal{Q}_{L})_{kl'ij}
 \right]
 \notag \\ &~~~
 + \frac{1}{2} \delta \overline{Z}^L_{k'k} (C^{(3)}_{\ell q})_{ijkl} (\mathcal{Q}_{L})_{k'lij}
 + \frac{1}{2} \delta Z^L_{ll'} (C^{(3)}_{\ell q})_{ijkl} (\mathcal{Q}_{L})_{kl'ij}
 \notag \\ &~~~
 + \left[ \frac{1}{2} \delta \overline{Z}^{eL}_{ii} + \frac{1}{2} \delta Z^{eL}_{jj} \right]
 (C^{(3)}_{\ell q})_{ijkl}
 (\mathcal{Q}_{L})_{klij}
 \notag \\ &=
 \frac{\alpha}{\pi s_W^2} \lambda_{m'}^{k'k} \lambda_{n'}^{ll'} 
 \left[ - J_2(x_{m'},x_{n'}) + \overline{K}_{k'k}(x_{m'},\mu) + K_{ll'}(x_{n'},\mu) \right] 
 (C^{(3)}_{\ell q})_{ijkl}
 (\mathcal{Q}_{L})_{k'l'ij}.
 \label{eq:C3lq}
\end{align}
Here, $\Xi_1$ and $\Xi_0$ depend on the gauge parameter.
The former, which is proportional to the quark mass, is canceled among the radiative corrections related to the quarks.
On the other hand, the latter cancellation is achieved by additionally taking account of those related to the leptons. 
Consequently, the result becomes independent of the gauge parameter. 
Finally, the matching formulae become
\begin{align}
 (C_9)^{\rm EW}_{ijkl} &= - (C_{10})^{\rm EW}_{ijkl}
 \notag \\ &=
 \frac{v^2}{s_W^2}\lambda_{m'}^{ii'} \lambda_{n'}^{j'j} 
 \left[ - J_2(x_{m'},x_{n'}) + \overline{K}_{ii'}(x_{m'},\mu) + K_{j'j}(x_{n'},\mu) \right] (C^{(3)}_{\ell q})_{kli'j'}
 \notag \\ &=
 \frac{v^2}{s_W^2} \lambda_{m'}^{ii'} \lambda_{n'}^{j'j} \, (C^{(3)}_{\ell q})_{kli'j'} \times
 \begin{cases}
 I_3(x_{m'},x_{n'},\mu),
 & (i' = i, j' = j) \\
 \left[ -J_2(x_{m'},x_{n'}) + K_1(x_{m'},\mu) \right],
 & (i' < i, j' = j) \\
 \left[ -J_2(x_{m'},x_{n'}) + K_2(x_{m'},\mu) \right],
 & (i' > i, j' = j) \\
 \left[ -J_2(x_{m'},x_{n'}) + K_1(x_{n'},\mu) \right],
 & (i' = i, j' < j) \\
 \left[ -J_2(x_{m'},x_{n'}) + K_2(x_{n'},\mu) \right],
 & (i' = i, j' > j) \\
 -J_2(x_{m'},x_{n'}),
 & (i' \neq i, j' \neq j) 
 \end{cases}\label{eq:c3lqloop}
\end{align}
where the result is summed over the primed indices.
Here, the loop function is defined as
\begin{align}
 I_3(x,y,\mu) &= -J_2(x,y) + K_0(x,\mu) + K_0(y,\mu),
\end{align}
which satisfies
\begin{align}
 I_3(x,\mu) &\equiv I_3(x,x,\mu) =
 -\frac{x}{16} \left[ 
   \ln\frac{\mu^2}{M_W^2}
 + \frac{7x-1}{2(x-1)} 
 - \frac{x^2-2x+4}{(x-1)^2} \ln x 
 \right].
\end{align}

Let us also consider the operator in the quark-flavor universal case,
\begin{align}
 (C^{(3)}_{\ell q})_{ijkl} &= 
 (C^{(3)}_{\ell q})_{ij} \delta_{kl}.
\end{align}
The matching formulae become
\begin{align}
 (C_9)^{\rm EW}_{ijkl} &= - (C_{10})^{\rm EW}_{ijkl}
 = \frac{v^2}{s_W^2} \lambda_{m'}^{ij} I_3(x_{m'},\mu) (C^{(3)}_{\ell q})_{kl},
\end{align}
where the result is summed over $m'$.
Since the top-quark contribution is dominant, the result is consistent with those in Ref.~\cite{Hurth:2019ula} by selecting $m'=3$.

\subsection{$(C_{Hu})_{ij}$}

Let us next consider the effective operators which include the Higgs fields.
First of all, the effective Lagrangian of $(C_{Hu})_{ij}$ is given by 
\begin{align}
 \mathcal{L}_{\rm eff} &= 
 (C_{Hu})_{ij}
 (H^\dagger i\overleftrightarrow{D_\mu} H) (\overline{u}_{i} \gamma^\mu P_R u_{j}),
\end{align}
which contributes to $d_i\to d_j \ell\ell$ by exchanging the $Z$ boson between the operator and the final-state leptons as in Fig.~\ref{fig:diagrams}(c). 
From Eq.~\eqref{eq:ve1}, the vertex correction to $u_R$ provides
\begin{align}
 \delta\mathcal{L}_{\rm eff} &= 
 \frac{\alpha}{\pi s_W^2}
 V^*_{m'k'} V_{n'l'} \,
 I_1(x_{m'},x_{n'},\mu) \,
 (C_{Hu})_{m'n'}
 (H^\dagger i\overleftrightarrow{D_\mu} H)
 (\overline{d}_{k'} \gamma^{\mu} P_L d_{l'}).
 \label{eq:CHu1}
\end{align}
For the effective operators including the Higgs fields, let us focus on the radiative corrections which change the quark flavors by the $W$-boson interactions.
Instead, if we consider the full EW radiative corrections, we additionally need to consider other corrections to the SMEFT vertex, vacuum polarizations, lepton self-energy corrections, and vertex corrections to the leptons. 
However, none of them change the quark flavors, and we checked that the gauge parameters are canceled among them. 
Thus, Eq.~\eqref{eq:CHu1} is sufficient to derive the flavor-changing matching conditions and show their gauge invariance. 
Consequently, the matching formulae are obtained as
\begin{align}
 (C_{9})_{ijkl}^{\rm EW} &= (-1+4s_W^2) (C_{10})_{ijkl}^{\rm EW}
 \notag \\ &=
 (-1+4s_W^2) \,
 \frac{v^2}{s_W^2} 
 V^*_{m'i} V_{n'j} \,
 I_1(x_{m'},x_{n'},\mu) \,
 (C_{Hu})_{m'n'} \delta_{kl},
\end{align}
where the result is summed over the primed indices.
It is noticed that the lepton flavor is universal, {\it i.e.}, $k=l$, because it is generated by the $Z$-boson interaction.

In particular, when the operator is universal for the quark flavor, 
\begin{align}
 (C_{Hu})_{ij} &= (C_{Hu})\, \delta_{ij},
\end{align}
the results are written as  
\begin{align}
 (C_{9})_{ijkl}^{\rm EW} &= (-1+4s_W^2) (C_{10})_{ijkl}^{\rm EW}
 \notag \\ &= 
 (-1+4s_W^2) \,
 \frac{v^2}{s_W^2} \lambda_{m'}^{ij} \,
 I_1(x_{m'},\mu) \,
 (C_{Hu})\, \delta_{kl}.
\end{align}
It is noticed that, by selecting $m'=3$, our result of $C_{10}$ is consistent with the result in Ref.~\cite{Hurth:2019ula}, but an overall sign of $C_{9}$ is opposite.

\subsection{$(C^{(1)}_{Hq})_{ij}$}

The effective operator $(\mathcal{O}^{(1)}_{Hq})_{ij}$ includes the left-handed quarks as well as the Higgs fields.
In the down basis, we obtain
\begin{align}
 \mathcal{L}_{\rm eff} &= 
 (C^{(1)}_{Hq})_{kl}
 (H^\dagger i\overleftrightarrow{D_\mu} H) (\overline{q}_{k} \gamma^\mu q_{l})
 \notag \\ &= 
 (C^{(1)}_{Hq})_{kl}
 (H^\dagger i\overleftrightarrow{D_\mu} H)
 \left[ 
 V_{m'k} V^*_{n'l} (\overline{u}_{m'} \gamma^{\mu} P_L u_{n'})
 + (\overline{d}_k \gamma^{\mu} P_L d_l)
 \right].
\end{align}
Similarly to the case of $(C_{Hu})_{ij}$, Eqs.~\eqref{eq:zdL}, \eqref{eq:zdbL}, and \eqref{eq:ve2} are enough to derive the flavor-changing corrections, which are assembled as\footnote{
 See Appendix \ref{sec:nonfl} for the gauge invariance of the flavor-unchanged contributions.
 \label{footnote:gaugeinv}
}
\begin{align}
 \delta\mathcal{L}_{\rm eff} &= 
 \frac{\alpha}{\pi s_W^2}
 \lambda_{m'}^{k'k} \lambda_{n'}^{ll'} 
 \big[ J_2(x_{m'},x_{n'}) + \Xi_1(x_{m'}) + \Xi_1(x_{n'}) + \Xi_0(\mu) \big]
 (C^{(1)}_{Hq})_{kl}
 (H^\dagger i\overleftrightarrow{D_\mu} H)
 (\overline{d}_{k'} \gamma^{\mu} P_L d_{l'}) 
 \notag \\ &~~~
 + \frac{1}{2} \delta \overline{Z}^L_{k'k} 
 (C^{(1)}_{Hq})_{kl}
 (H^\dagger i\overleftrightarrow{D_\mu} H)
 (\overline{d}_{k'} \gamma^{\mu} P_L d_{l})
 + \frac{1}{2} \delta Z^L_{ll'} 
 (C^{(1)}_{Hq})_{kl}
 (H^\dagger i\overleftrightarrow{D_\mu} H)
 (\overline{d}_{k} \gamma^{\mu} P_L d_{l'})
 \notag \\ &=
 \frac{\alpha}{\pi s_W^2} \lambda_{m'}^{k'k} \lambda_{n'}^{ll'} 
 \big[ J_2(x_{m'},x_{n'}) + \overline{K}_{k'k}(x_{m'},\mu) + K_{ll'}(x_{n'},\mu) \big] 
 (C^{(1)}_{Hq})_{kl}
 (H^\dagger i\overleftrightarrow{D_\mu} H)
 (\overline{d}_{k'} \gamma^{\mu} P_L d_{l'}).\label{eq:c1Hq}
\end{align}
It is found that the gauge parameter is canceled.
Then, the matching formulae become
\begin{align}
 (C_{9})_{ijkl}^{\rm EW} &= (-1+4s_W^2) (C_{10})_{ijkl}^{\rm EW}
 \notag \\ &=
 (-1+4s_W^2) \,
 \frac{v^2}{s_W^2} \lambda_{m'}^{ii'} \lambda_{n'}^{j'j} 
 \left[ J_2(x_{m'},x_{n'}) + \overline{K}_{ii'}(x_{m'},\mu) + K_{j'j}(x_{n'},\mu) \right] 
 (C^{(1)}_{Hq})_{i'j'} \, \delta_{kl}
 \notag \\ &=
 (-1+4s_W^2) \,
 \frac{v^2}{s_W^2} \lambda_{m'}^{ii'} \lambda_{n'}^{j'j} \,
 (C^{(1)}_{Hq})_{i'j'} \, \delta_{kl}
 \notag \\ &\qquad\qquad\qquad \times
 \begin{cases}
 \left[ J_2(x_{m'},x_{n'}) + K_1(x_{m'},\mu) \right],
 & (i' < i, j' = j) \\
 \left[ J_2(x_{m'},x_{n'}) + K_2(x_{m'},\mu) \right],
 & (i' > i, j' = j) \\
 \left[ J_2(x_{m'},x_{n'}) + K_1(x_{n'},\mu) \right],
 & (i' = i, j' < j) \\
 \left[ J_2(x_{m'},x_{n'}) + K_2(x_{n'},\mu) \right],
 & (i' = i, j' > j) \\
 J_2(x_{m'},x_{n'}),
 & (i' \neq i, j' \neq j) 
 \end{cases}
\end{align}
where the result is summed over the primed indices.

In the quark-flavor universal case,
\begin{align}
 (C^{(1)}_{Hq})_{ij} &= (C^{(1)}_{Hq})\, \delta_{ij},
\end{align}
the above results become
\begin{align}
 (C_{9})_{ijkl}^{\rm EW} 
 = (-1+4s_W^2) (C_{10})_{ijkl}^{\rm EW}
 = - (-1+4s_W^2) \, 
 \frac{v^2}{s_W^2} \lambda_{m'}^{ij} \,
 I_1(x_{m'},\mu) \,
 (C^{(1)}_{Hq})\, \delta_{kl}.
\end{align}
It is noticed that, by selecting $m'=3$, our results are consistent with those in Ref.~\cite{Hurth:2019ula} except for an overall sign of $C_{9}$.

\subsection{$(C^{(3)}_{Hq})_{ij}$}

In the down basis, the effective operator $(\mathcal{O}^{(3)}_{Hq})_{ij}$ is expanded as
\begin{align}
 \mathcal{L}_{\rm eff} &= 
 (C^{(3)}_{Hq})_{kl}
 (H^\dagger i\overleftrightarrow{D_\mu^a} H) (\overline{q}_{k} \gamma^\mu \tau^a q_{l})
 \notag \\ &= 
 \frac{gv^2}{2c_W} 
 (C^{(3)}_{Hq})_{kl}\,
 Z_\mu
 \left[ 
 V_{m'k} V^*_{n'l} (\overline{u}_{m'} \gamma^{\mu} P_L u_{n'})
 - (\overline{d}_k \gamma^{\mu} P_L d_l)
 \right]
 \notag \\ &\phantom{=} + 
 \frac{gv^2}{\sqrt{2}} 
 (C^{(3)}_{Hq})_{kl}
 \Big[V_{m'k} \overline{u}_{m'} \gamma^\mu P_L d_{l} W_\mu + V_{n'l}^* \overline{d}_k \gamma^\mu P_L u_{n'} W_\mu^\dagger \Big]
 + \cdots,
\end{align}
where $c_W$ is the cosine of the Weinberg angle. 
The Feynman rules in the $R_{\xi}$ gauge are found, {\it e.g.}, in Ref.~\cite{Dedes:2017zog}. 
In order to derive the EW radiative corrections which change the quark flavors by the CKM matrix, we need to consider Eqs.~\eqref{eq:zdL}, \eqref{eq:zdbL}, \eqref{eq:ve2}, \eqref{eq:box}, \eqref{eq:phpen}, and \eqref{eq:zphpen}, which are assembled as
\begin{align}
 \delta\mathcal{L}_{\rm eff} &= 
 \delta\mathcal{L}^{(1)} + \delta\mathcal{L}^{(2)} + \delta\mathcal{L}^{(3)}.
\end{align}
In the right-hand side, the first term is from Eq.~\eqref{eq:ve2}, {\it i.e.}, the diagram in which the $Z$ boson is exchanged between the SMEFT operator and the leptons, and the result is
\begin{align}
 \delta\mathcal{L}^{(1)} &=
 -\frac{\alpha}{\pi s_W^2}
 \lambda_{m'}^{k'k} \lambda_{n'}^{ll'} 
 \big[ J_2(x_{m'},x_{n'}) + \Xi_1(x_{m'}) + \Xi_1(x_{n'}) \big]
 (C^{(3)}_{Hq})_{kl} (\mathcal{Q}_{Z})_{k'l'ii}.\label{eq:dl1}
\end{align}
The second term assembles the other contributions which depends on $\lambda_{m'}^{k'k}$ as
\begin{align}
 \delta\mathcal{L}^{(2)} &=
 - \frac{\alpha}{\pi s_W^2}
 \lambda_{m'}^{k'k} 
 \left[ B(x_{m'}) + \Xi_2(x_{m'}) \right] 
 (C^{(3)}_{Hq})_{kl} (\mathcal{Q}_{L})_{k'lii} 
 \\ &~~
 - \frac{\alpha}{\pi s_W^2}
 \lambda_{m'}^{k'k} 
 \left[ s_W^2 D(x_{m'}) - 2 s_W^2 \Xi_2(x_{m'}) \right] 
 (C^{(3)}_{Hq})_{kl} \left[ (\mathcal{Q}_{L})_{k'lii} + (\mathcal{Q}_{R})_{k'lii} \right]
 \notag \\ &~~
 - \frac{\alpha}{\pi s_W^2}
 \lambda_{m'}^{k'k} 
 \left[
   J_3(x_{m'},\mu) + \Xi_2(x_{m'}) - 3 \Xi_1(x_{m'})
   - \frac{2}{3} s_W^2 \left[ K_1(x_{m'},\mu) - \Xi_1(x_{m'}) \right]
 \right]
 (C^{(3)}_{Hq})_{kl} (\mathcal{Q}_{Z})_{k'lii} 
 \notag \\ &~~
 + \frac{1}{2} \delta \overline{Z}^L_{k'k} (C^{(3)}_{Hq})_{kl} (\mathcal{Q}_{Z})_{k'lii}
 + \frac{\alpha}{\pi s_W^2}
 \lambda_{m'}^{k'k} 
 \left[ 1 - \frac{2}{3}s_W^2 \right]
 \left[ K_1(x_{m'},\mu) - \Xi_1(x_{m'}) \right] 
 (C^{(3)}_{Hq})_{kl} (\mathcal{Q}_{Z})_{k'lii},
 \notag
\end{align}
and the third term corresponds to those proportional to $\lambda_{n'}^{ll'}$ as
\begin{align}
 \delta\mathcal{L}^{(3)} &=
 - \frac{\alpha}{\pi s_W^2}
 \lambda_{n'}^{ll'} 
 \left[ B(x_{n'}) + \Xi_2(x_{n'}) \right] 
 (C^{(3)}_{Hq})_{kl} (\mathcal{Q}_{L})_{kl'ii} 
 \\ &~~
 - \frac{\alpha}{\pi s_W^2}
 \lambda_{n'}^{ll'} 
 \left[ s_W^2 D(x_{n'}) - 2 s_W^2 \Xi_2(x_{n'}) \right] 
 (C^{(3)}_{Hq})_{kl} \left[ (\mathcal{Q}_{L})_{kl'ii} + (\mathcal{Q}_{R})_{kl'ii} \right]
 \notag \\ &~~
 - \frac{\alpha}{\pi s_W^2}
 \lambda_{n'}^{ll'} 
 \left[
   J_3(x_{n'},\mu) + \Xi_2(x_{n'}) - 3 \Xi_1(x_{n'})
   - \frac{2}{3} s_W^2 \left[ K_1(x_{n'},\mu) - \Xi_1(x_{n'}) \right]
 \right]
 (C^{(3)}_{Hq})_{kl} (\mathcal{Q}_{Z})_{kl'ii} 
 \notag \\ &~~
 + \frac{1}{2} \delta Z^L_{ll'} (C^{(3)}_{Hq})_{kl} (\mathcal{Q}_{Z})_{kl'ii}
 + \frac{\alpha}{\pi s_W^2}
 \lambda_{n'}^{ll'} 
 \left[ 1 - \frac{2}{3}s_W^2 \right]
 \left[ K_1(x_{n'},\mu) - \Xi_1(x_{n'}) \right] 
 (C^{(3)}_{Hq})_{kl} (\mathcal{Q}_{Z})_{kl'ii}.
 \notag
\end{align}
Here, the summation over the primed indices are understood.
Consequently, we obtain
\begin{align}
 \delta\mathcal{L}_{\rm eff} &= 
 - \frac{\alpha}{\pi s_W^2} \lambda_{m'}^{k'k} \lambda_{n'}^{ll'} 
 \Big[
 (-1+2s_W^2) \,
 J_2(x_{m'},x_{n'}) \,
 \notag \\ &\qquad\qquad\qquad\quad
 +
 \Big\{
 (-1+2s_W^2) \left[ J_3(x_{m'},\mu) - K_1(x_{m'},\mu) - \overline{K}_{k'k}(x_{m'},\mu) \right]
 \notag \\ &\qquad\qquad\qquad\quad
 + \left[ B(x_{m'}) + s_W^2 D(x_{m'}) \right] 
 + (x_{m'} \to x_{n'}, \overline{K}_{k'k} \to K_{ll'}) 
 \Big\} 
 \Big] \,
 (C^{(3)}_{Hq})_{kl} 
 (\mathcal{O}_{L})_{k'l'ii} 
 \notag \\ &~~~
 - \frac{\alpha}{\pi s_W^2} \lambda_{m'}^{k'k} \lambda_{n'}^{ll'}
 \Big[\, 
 2 s_W^2 J_2(x_{m'},x_{n'})
 \notag \\ &\qquad\qquad\qquad\quad
 +
 s_W^2 \Big\{\,
 2 \left[ J_3(x_{m'},\mu) - K_1(x_{m'},\mu) - \overline{K}_{k'k}(x_{m'},\mu) \right]
 + D(x_{m'}) 
 \notag \\ &\qquad\qquad\qquad\quad
 + (x_{m'} \to x_{n'}, \overline{K}_{k'k} \to K_{ll'}) \Big\} 
 \Big] \,
 (C^{(3)}_{Hq})_{kl} 
 (\mathcal{O}_{R})_{k'l'ii}.
 \label{eq:c3hq}
\end{align}
It is noticed that the gauge parameters via $\Xi_1$ and $\Xi_2$ are canceled completely. 
In the above results, we omit the contributions which do not change the quark flavors, {\it e.g.}, those via $\Xi_0$.
The gauge dependencies in such contributions are also canceled by including additional radiative corrections (see Appendix \ref{sec:nonfl}).

As a result, the matching formulae are determined as
\begin{align}
 \label{eq:9chq3}
 (C_{9})_{ijkl}^{\rm EW} &= 
 -\frac{v^2}{s_W^2} \lambda_{m'}^{ii'} \lambda_{n'}^{j'j} 
 \Big\{ 
 (-1+4s_W^2) \left[ J_2(x_{m'},x_{n'})
 + J_3(x_{m'},\mu) + J_3(x_{n'},\mu)
 \right.
 \\ &\qquad\qquad\qquad\quad
 \left.
 - K_1(x_{m'},\mu) - K_1(x_{n'},\mu)
 - \overline{K}_{ii'}(x_{m'},\mu) - K_{j'j}(x_{n'},\mu) \right] 
 \notag \\ &\qquad\qquad\qquad~
 + \left[ B(x_{m'}) + B(x_{n'}) + 2 s_W^2 D(x_{m'}) + 2 s_W^2 D(x_{n'}) \right]
 \Big\} \,
 (C^{(3)}_{Hq})_{i'j'} \, \delta_{kl},
 \notag \\
 \label{eq:10chq3}
 (C_{10})_{ijkl}^{\rm EW} &= 
 -\frac{v^2}{s_W^2} \lambda_{m'}^{ii'} \lambda_{n'}^{j'j} 
 \Big[ \,
 J_2(x_{m'},x_{n'})
 + J_3(x_{m'},\mu) + J_3(x_{n'},\mu)
 \\ &\qquad\qquad\qquad~
 - K_1(x_{m'},\mu) - K_1(x_{n'},\mu)
 - \overline{K}_{ii'}(x_{m'},\mu) - K_{j'j}(x_{n'},\mu) 
 \notag \\ &\qquad\qquad\qquad~
 - B(x_{m'}) - B(x_{n'})
 \, \Big] \,
 (C^{(3)}_{Hq})_{i'j'} \, \delta_{kl},
 \notag
\end{align}
where the results are summed over the primed indices.
It is noticed again that the loop functions depend on the quark indices, $i,i',j,j'$, via the wave-function renormalization.

In order to compare the above results with those in Ref.~\cite{Hurth:2019ula}, let us consider the operator in the flavor-universal case, 
\begin{align}
 (C^{(3)}_{Hq})_{ij} &= (C^{(3)}_{Hq})\, \delta_{ij}.
\end{align}
Then, the formulae become
\begin{align}
 (C_{9})_{ijkl}^{\rm EW} &= 
 -\frac{2 v^2}{s_W^2} \lambda_{m'}^{ij} 
 \bigg\{ \,
 (-1+4s_W^2) 
 \left[ 
 \frac{1}{2} J_2(x_{m'}) 
 + J_3(x_{m'},\mu) 
 - K_1(x_{m'},\mu) 
 - K_0(x_{m'},\mu) 
 \right] 
 \notag \\ &\qquad\qquad~~~
 + \left[ 
 B(x_{m'}) 
 + 2 s_W^2 D(x_{m'}) 
 \right]
 \bigg\} \,
 (C^{(3)}_{Hq})\, \delta_{kl},
 \\
 (C_{10})_{ijkl}^{\rm EW} &= 
 - \frac{2 v^2}{s_W^2} \lambda_{m'}^{ij} 
 \bigg[
 \frac{1}{2} J_2(x_{m'})
 + J_3(x_{m'},\mu)
 \notag \\ &\qquad\qquad~~~
 - K_1(x_{m'},\mu)
 - K_0(x_{m'},\mu)
 - B(x_{m'})
 \bigg]
 (C^{(3)}_{Hq}) \delta_{kl}.
\end{align}
In Ref.~\cite{Hurth:2019ula}, the corresponding results are provided in the 't Hooft-Feynman gauge, $\xi_W = 1$.
Their loop functions are related to ours as
\begin{align}
 I^{Hq3}(x) &= -\frac{1}{2} J_2(x) - J_3(x,\mu) + K_1(x,\mu) + K_0(x,\mu) - \Xi_2(x)\big|_{\xi_W=1}, \\
 B_0(x) &= B(x) +\Xi_2(x)\big|_{\xi_W=1}, \\
 \frac{1}{2} D_0(x) &= D(x) - 2 \, \Xi_2(x)\big|_{\xi_W=1},
\end{align}
where the left-hand side is defined in Ref.~\cite{Hurth:2019ula}.
It is noticed that, by selecting $m'=3$, our result is consistent with theirs if a sign of $I^{Hq3}(x)$ is fixed as
\begin{align}
 I^{Hq3}(x) = \frac{x}{32}
 \left[ 7 \ln \frac{\mu^2}{M_W^2}-\frac{x+33}{2(1-x)}-\frac{7x^2-2x+12}{(1-x)^2}\ln x \right].
\end{align}

\section{Application}

As an application of the matching formulae provided in the previous section, let us study the $b \to s \mu^+\mu^-$ transitions.
Based on the latest experimental results including that on $B^+ \to K^{*+}\mu^+\mu^-$, the global-fit analysis implies that the NP contributions are \cite{Ciuchini:2020gvn}
\begin{align}
 (C^{(1,3)}_{\ell q})_{2223} &= 
 \frac{1}{\Lambda^2}(0.77 \pm 0.13), ~~~\mbox{(1D)}
 \label{eq:result1_1D} \\
 \left[ (C^{(1,3)}_{\ell q})_{2223}, (C_{q e})_{2322} \right] &= 
 \frac{1}{\Lambda^2} \left[ 0.80 \pm 0.18, 0.05 \pm 0.30 \right], ~~~\mbox{(2D)}
 \label{eq:result1_2D}
\end{align}
from the 1-dimensional (1D) and 2-dimensional (2D) fits, respectively, with the normalization factor, $\Lambda=30\TeV$.\footnote{
In Ref.~\cite{Ciuchini:2020gvn}, it is not clear whether the RG corrections above the EW scale are taken into account for these SMEFT operators. 
We discard them in the following analysis.
}
Here, a data-driven method is used to estimate non-factorizable 
hadronic contributions associated with the charm loop. 
On the other hand, if the hadronic contributions are determined by the $q^2$ extrapolation of the light-cone sum rules estimate~\cite{Khodjamirian:2010vf}\footnote{
Recently, updated results on the hadronic contributions have been provided in Ref.~\cite{Gubernari:2020eft}.
}, the analysis provides
\begin{align}
 (C^{(1,3)}_{\ell q})_{2223} &= 
 \frac{1}{\Lambda^2}(0.92 \pm 0.12), ~~~\mbox{(1D)} 
 \label{eq:result2_1D} \\
 \left[ (C^{(1,3)}_{\ell q})_{2223}, (C_{q e})_{2322} \right] &= 
 \frac{1}{\Lambda^2} \left[ 1.03 \pm 0.12, 0.71 \pm 0.13 \right], ~~~\mbox{(2D)}
 \label{eq:result2_2D} 
\end{align}
for $\Lambda=30\TeV$.
Note that they contribute to $b \to s \mu^+\mu^-$ at the tree level.

Let us consider the following flavor-conserving operators,
\begin{align}
 \mathcal{O}_i = \{ 
 (\mathcal{O}_{e u})_{2233}, 
 (\mathcal{O}_{\ell u})_{2233}, 
 (\mathcal{O}_{q e})_{3322}, 
 (\mathcal{O}^{(1)}_{\ell q})_{2233}, 
 (\mathcal{O}^{(3)}_{\ell q})_{2233} \},
 \label{eq:operators}
\end{align}
at the EWSB scale, $\mu = M_W$.\footnote{For investigating other SMEFT operators we need detailed analyses and retain for future works.}
They do not change the quark flavors at the tree level, but induce $b \to s \mu^+\mu^-$ via EW radiative corrections.
The LEFT operators, $Q_9$ and $Q_{10}$, receive contributions from them in very similar to the above tree-level contributions from $(\mathcal{O}^{(1,3)}_{\ell q})_{2223}$ and $(\mathcal{O}_{q e})_{2322}$.
In fact, the LEFT Wilson coefficients, $(C_9)_{2322}$ and $(C_{10})_{2322}$, are determined from Eqs.~\eqref{eq:result1_1D}\text{--}\eqref{eq:result2_2D} by using the tree-level matching conditions, Eqs.~\eqref{eq:tree1} and \eqref{eq:tree2}.
These LEFT results can be reproduced by the above operators \eqref{eq:operators} via the EW radiative corrections, Eqs.~\eqref{eq:euloop}, \eqref{eq:cluloop}, \eqref{eq:ResultCqe}, \eqref{eq:c1lqloop}, and \eqref{eq:c3lqloop}.
For example, when one considers the contributions to $(C_{9,10})_{ijkl}$ with $\{i,j,k,l\} = \{2,3,2,2\}$ from $(C^{(1)}_{\ell q})_{kli'j'}$ with $\{k,l,i',j'\}=\{2,2,3,3\}$, the one-loop matching formula is provided by the third relation in Eq.~\eqref{eq:c1lqloop} because of $i < i'$ and $j = j'$.
Since the contributions satisfy $(C_9)_{2322}=-(C_{10})_{2322}$, one notices that Eq.~\eqref{eq:result1_1D} or \eqref{eq:result2_1D} can be recast to derive the result according to Eqs.~\eqref{eq:tree1} and \eqref{eq:tree2}.
Our result is summarized in Table~\ref{tab:fit}, where ``Analysis1'' corresponds to Eqs.~\eqref{eq:result1_1D} and \eqref{eq:result1_2D}, and ``Analysis2'' to \eqref{eq:result2_1D} and \eqref{eq:result2_2D}.
For the input parameters we used $G_F=1.1663787\times 10^{-5}\,{\rm GeV}^{-2}$, $\sin^2 \theta_W=0.23121$, $\alpha (M_Z)=1/127.952$, $M_W=80.379\,{\rm  GeV}$, and $m_t=172.76\,{\rm GeV}$.
Also the Wolfenstein parameters are $\lambda=0.22650$, $A=0.790$, $\bar{\rho}=0.141$ and $\bar{\eta}=0.357$.
In the numerical analysis, we neglected uncertainties for these parameters because they are sufficiently small. 
As a result, it is found that the NP scale is typically around $0.5$--$1\TeV$.\footnote{
Experimental constraints on these operators have been studied, {\it e.g.}, in Ref.~\cite{Coy:2019rfr}.
}

It is noticed that we set $\mu = M_W$ to derive Table~\ref{tab:fit}, and thus, the logarithmic term in the loop functions of the EW radiative corrections vanish, but the contributions are provided by the finite terms. 
This is contrasted to the analyses in Refs.~\cite{Coy:2019rfr,Aebischer:2020lsx,Alasfar:2020mne}, where the contributions are generated by the RG corrections, i.e., the logarithmic term by assuming that the input scale of the SMEFT operators are much larger than the EW scale and discarding the finite terms. 
Consequently, our results differ from them by factors of $O(1)$.

\begin{table}[t]
\centering
\begin{tabular}{|c||c|c|}
\hline
& Analysis1 
& Analysis2\\ \hline \hline
$ c_{\ell u} $ & 
$ \phantom{-} 4.5 \pm 0.8 $ & 
$ \phantom{-} 5.3 \pm 0.7 $ \\ \hline
$ c^{(1)}_{\ell q} $ & 
$ -2.3 \pm 0.4 $ &
$ -2.7 \pm 0.4 $ \\ \hline
$ c^{(3)}_{\ell q} $ & 
$ \phantom{-} 7.4 \pm 1.3 $ &
$ \phantom{-} 8.9 \pm 1.2 $ \\ \hline \hline
$ ( c_{\ell u}, c_{e u} ) $ & 
$ (\phantom{-} 4.6 \pm 1.0, \phantom{-} 0.3 \pm 1.7) $ &
$ (\phantom{-} 6.0 \pm 0.7, \phantom{-} 4.1 \pm 0.8) $ \\ \hline
$ ( c^{(1)}_{\ell q}, c_{q e} ) $ & 
$ (-2.4 \pm 0.5, -0.1 \pm 0.9) $ &
$ (-3.0 \pm 0.4, -2.1 \pm 0.4) $ \\ \hline
$ ( c^{(3)}_{\ell q}, c_{q e} ) $ & 
$ (\phantom{-} 7.7 \pm 1.7, -0.1 \pm 0.9) $ &
$ (\phantom{-} 9.9 \pm 1.2, -2.1 \pm 0.4) $ \\
\hline
\end{tabular}
\caption{Values of Wilson coefficients, $C_i = c_i/(1\TeV)^2$, for the flavor-conserving SMEFT operators, \eqref{eq:operators}. 
Here, ``Analysis1'' is derived by recasting \eqref{eq:result1_1D} and \eqref{eq:result1_2D}, and ``Analysis2'' from \eqref{eq:result2_1D} and \eqref{eq:result2_2D}.
The upper (lower) group is obtained by the 1D (2D) global fits. }
\label{tab:fit}
\end{table}

\section{Conclusions and discussion}

In this article, the EW radiative corrections to the SMEFT operators were revisited for the current $b \to s\ell\ell$ results.
We provided the matching condition formulae between the SMEFT and LEFT at the one-loop level. 
We did not assumed any flavor symmetry {\it a priori}, and thus, flavor structures of the operators are general.
The gauge-parameter cancellations were shown explicitly by performing the calculations in the $R_\xi$ gauge. 
Besides, the on-shell conditions were applied especially in dealing with the quark-flavor mixings appropriately. 
Consequently, it was noticed that the results depend on the flavor structure via the wave-function renormalization.

In light of the recent results on $B^+ \to K^{*+}\mu^+\mu^-$ as well as $B^0 \to K^{*0}\mu^+\mu^-$, we have also studied an implication of our formulae on the current $b \to s\ell\ell$ anomalies. 
If they are induced by flavor-conserving NP contributions where the flavor transitions are triggered by the EW radiative corrections, the NP scale is estimated to be around $0.5$--$1\TeV$.

In this article, we have focused on the $b \to s\ell\ell$ transitions. 
It is straightforward to apply our analyses to determine EW radiative corrections to other SMEFT operators, {\it e.g.}, the four-quark operators relevant for meson mixings. 
For instance, loop functions of the $Z$-mediated NP contributions shall depend on the flavor structure.
The details will be discussed in our future works.

\section{Acknowledgements}
This work is supported in part by the Grant-in-Aid for
Scientific Research B (No.16H03991 [ME]), 
Early-Career Scientists (No.16K17681 [ME]) and 
Scientific Research C (No.17K05429 [SM]). 
The diagrams in Fig.~\ref{fig:diagrams} are drawn with {\tt TikZ-Feynman} \cite{Ellis:2016jkw}.

\appendix
\section{Flavor-unchanged contribution for $C^{(1,3)}_{Hq}$}
\label{sec:nonfl}
In this section, we present the $W$-boson loops contributions which do not change the quark flavors by the CKM matrix for the SMEFT operators $\mathcal{O}^{(1,3)}_{Hq}$.
In particular, we will show the gauge-parameter cancellations by including the diagrams which are omitted in Sec.~\ref{sec:EWoneloop}.
For $C^{(1)}_{Hq}$, the one-loop contributions are shown as
\begin{align}
\delta \mathcal{L}_{\rm eff} = 
\delta\mathcal{L}^{\rm self}_{Hq1} 
+ \delta \mathcal{L}^{\rm vac}_{Hq1} 
+ \delta\mathcal{L}^{\rm vert}_{Hq1}.
\end{align}
In the right-hand side, the first term arises from the self-energy corrections \eqref{eq:zdL}, \eqref{eq:zdbL}, and \eqref{eq:zeL} as
\begin{align}
\delta \mathcal{L}^{\rm self}_{Hq1} &=
\left( \frac{1}{2}\delta \bar{Z}^L_{kk} + \frac{1}{2}\delta {Z}^L_{ll} \right)
(C^{(1)}_{Hq})_{kl} (\mathcal{Q}_Z)_{klii}
+ \left( \frac{1}{2}\delta \bar{Z}^{eL}_{ii} + \frac{1}{2}\delta {Z}^{eL}_{ii} \right)
(-1+2 s_W^2)(C^{(1)}_{Hq})_{kl} (\mathcal{Q}_L)_{klii}.
\end{align}
The second term corresponds to the photon/$Z$ vacuum polarizations due to the $W$ (NG) boson loops,
\begin{align}
\delta \mathcal{L}^{\rm vac}_{Hq1} &=
\frac{\alpha}{\pi s_W^2}
(C^{(1)}_{Hq})_{kl} (\mathcal{Q}_Z)_{klii}
\notag \\ &~~
\times 
\frac{1}{8} (-1+2 s_W^2)
\left[
(3+\xi_W)\ln \frac{\mu^2}{M_W^2} 
+ \frac{1}{2}(5 + \xi_W)
-\frac{\xi_W+2}{\xi_W-1}\xi_W\ln \xi_W
\right]
\notag \\ &~~
- \frac{\alpha}{\pi s_W^2}
(C^{(1)}_{Hq})_{kl} \left[(\mathcal{Q}_L)_{klii}+(\mathcal{Q}_R)_{klii}
\right]
\notag \\ &~~
\times 
\frac{1}{12} s_W^2 \bigg\{
\left[ 34 -3\xi_W -3s_W^2 (11 -\xi_W) \right]
\ln \frac{\mu^2}{M_W^2}
\notag \\ &\qquad\qquad~
+ \frac{20 +47\xi_W +2\xi_W^2-9\xi_W^3}{6(\xi_W-1)^2}
+ \frac{1}{2} s_W^2 (7 +3 \xi_W) 
\notag \\ &\qquad\qquad~
- \frac{6 +6\xi_W +\xi_W^2 -3\xi_W^3 +3 s_W^2 (2-3\xi_W +\xi_W^3)}{(\xi_W-1)^3} \xi_W \ln \xi_W
\bigg\}.
\end{align}
The third term is from the vertex corrections to the photon/$Z$ penguin contributions.
In addition to \eqref{eq:ve2}, there are contributions in which the $W$-boson interactions are not accompanied by the CKM matrix, and thus, they do not depend on $\lambda^{kk}_{n'}$ or $\lambda^{ll}_{n'}$.
The result is given by
\begin{align}
\mathcal{L}^{\rm vert}_{Hq1} &=
\frac{\alpha}{\pi s_W^2} \lambda_{m'}^{kk} \lambda_{n'}^{ll}
\bigg[
 J_2(x_{m'},x_{n'})+\Xi_1(x_{m'})+\Xi_1(x_{n'})+\Xi_0 (\mu)
\bigg]
(C^{(1)}_{Hq})_{kl} (\mathcal{Q}_Z)_{klii}
\notag \\ &~~
- \frac{\alpha}{\pi s_W^2}(C_{Hq}^{(1)})_{kl}(\mathcal{Q}_L)_{klii}
\notag \\ &~~
\times
\frac{1}{8}
\bigg\{
\left[ (3 +2 \xi_W) -3 s_W^2 (1+\xi_W) \right] \ln \frac{\mu^2}{M_W^2}
\notag \\ &\qquad~~
+ \frac{2 +3\xi_W +s_W^2 (1 -5 \xi_W)}{2}
- \left[ 1+(2-3 s_W^2)\xi_W \right] \frac{\xi_W\ln \xi_W}{\xi_W-1} 
\bigg\}
\notag \\ &~~
- \frac{\alpha}{\pi s_W^2}
(C^{(1)}_{Hq})_{kl} (\mathcal{Q}_Z)_{klii} 
\times
\frac{1}{8} s_W^2 
\bigg[
(3 +\xi_W) \ln \frac{\mu^2}{M_W^2}
+ \frac{1}{2}(5+\xi_W) 
- \frac{\xi_W+2}{\xi_W-1}\xi_W\ln\xi_W
\bigg]
\notag \\ &~~
+ \frac{\alpha}{\pi s_W^2}
(C^{(1)}_{Hq})_{kl} \left[(\mathcal{Q}_L)_{klii}+(\mathcal{Q}_R)_{klii}\right]
\notag \\ &~~
\times
\frac{1}{12} s_W^2
\left[
\ln \frac{\mu^2}{M_W^2}
+ \frac{41+14\xi_W+5\xi_W^2}{6(\xi_W-1)^2}
- \frac{12 -3\xi_W +\xi_W^2}{(\xi_W-1)^3}\xi_W \ln \xi_W
\right]
.
\end{align}
Consequently, we obtain
\begin{align}
\delta \mathcal{L}_{\rm eff} &=
\frac{\alpha}{\pi s_W^2}
\lambda^{kk}_{m'}\lambda^{ll}_{n'}
(C_{Hq}^{(1)})_{kl} (\mathcal{Q}_L)_{klii}
\notag \\ &\times
\left\{
(-1+2s_W^2)
\left[ J_2(x_{m'},x_{n'})+K_0(x_{m'},\mu)+K_0(x_{n'},\mu) \right]
+ s_W^2 (-1+s_W^2) 
\left( \frac{7}{2} \ln\frac{\mu^2}{M_W^2} + \frac{1}{3} \right)
\right\}
\notag \\ &+
\frac{\alpha}{\pi s_W^2}
\lambda^{kk}_{m'}\lambda^{ll}_{n'}
(C_{Hq}^{(1)})_{kl} (\mathcal{Q}_R)_{klii}
\notag \\ &\times
\left\{
2 s_W^2
\left[ J_2(x_{m'},x_{n'})+K_0(x_{m'},\mu)+K_0(x_{n'},\mu) \right]
+ s_W^2 (-1+s_W^2) 
\left( \frac{7}{2} \ln\frac{\mu^2}{M_W^2} + \frac{1}{3} \right)
\right\}.
\end{align}
Thus, the gauge parameters in the $W$-boson loop contributions are canceled completely.

On the other hand, the flavor-unchanged contributions for $C^{(3)}_{Hq}$ from the $W$-boson loops are given by
\begin{align}
\delta\mathcal{L}_{\rm eff} =
\delta \mathcal{L}^{\rm self}_{Hq3}
+ \delta \mathcal{L}^{\rm vac}_{Hq3}
+ \delta \mathcal{L}^{\rm box+peng}_{Hq3}
+ \delta \mathcal{L}^{\rm vert}_{Hq3},
\end{align}
where the first term in the right-hand side arises from the self-energy corrections obtained from Eqs.~\eqref{eq:zdL}, \eqref{eq:zdbL}, \eqref{eq:zeL}, and those corresponding to Eq.~\eqref{eq:wtnKHQ3} for $i=j$.
The result becomes
\begin{align}
\delta \mathcal{L}^{\rm self}_{Hq3} &=
\left( \frac{1}{2}\delta \bar{Z}^L_{kk} + \frac{1}{2}\delta {Z}^L_{ll} \right)
(C^{(3)}_{Hq})_{kl} (\mathcal{Q}_Z)_{klii}
+ \left( \frac{1}{2}\delta \bar{Z}^{eL}_{ii} + \frac{1}{2}\delta {Z}^{eL}_{ii} \right)
(-1+2 s_W^2)
(C^{(3)}_{Hq})_{kl} (\mathcal{Q}_L)_{klii}
\notag \\ &~~
+ \frac{\alpha}{\pi s_W^2} \left[ \lambda^{kk}_{m'} + \lambda^{ll}_{m'} \right]
(C^{(3)}_{Hq})_{kl} (\mathcal{Q}_Z)_{klii}
\notag \\ &~~
\times \left( 1-\frac{2}{3}s_W^2 \right)
\bigg\{
K_1(x_{m'},\mu)-\Xi_1(x_{m'})
- \frac{1}{8} \left[ \xi_W \ln \frac{\mu^2}{M_W^2} - \frac{3}{2} +\xi_W(1 -\ln \xi_W) \right]
\bigg\}.
\end{align}
The second term is from the photon/$Z$ vacuum polarizations as
\begin{align}
\delta \mathcal{L}^{\rm vac}_{Hq3} &=
\frac{\alpha}{\pi s_W^2}
(C^{(3)}_{Hq})_{kl} (\mathcal{Q}_Z)_{klii}
\notag \\ &~~
\times 
\frac{1}{8} (-1+2 s_W^2)
\left[
(3+\xi_W)\ln \frac{\mu^2}{M_W^2} 
+ \frac{1}{2}(5 + \xi_W)
-\frac{\xi_W+2}{\xi_W-1}\xi_W \ln \xi_W
\right]
\notag \\ &~~
- \frac{\alpha}{\pi s_W^2}
(C^{(3)}_{Hq})_{kl} \left[(\mathcal{Q}_L)_{klii}+(\mathcal{Q}_R)_{klii}
\right]
\notag \\ &~~
\times 
\frac{1}{12} s_W^2 \bigg\{
\left[ 34 -3\xi_W -3s_W^2 (11 -\xi_W) \right]
\ln \frac{\mu^2}{M_W^2}
\notag \\ &\qquad\qquad~
+ \frac{20 +47\xi_W +2\xi_W^2-9\xi_W^3}{6(\xi_W-1)^2}
+ \frac{1}{2} s_W^2 (7 +3 \xi_W) 
\notag \\ &\qquad\qquad~
- \frac{6 +6\xi_W +\xi_W^2 -3\xi_W^3 +3 s_W^2 (2-3\xi_W +\xi_W^3)}{(\xi_W-1)^3} \xi_W \ln \xi_W
\bigg\}.
\end{align}
The third term corresponds to the box and penguin contributions discussed in Sec.~\ref{sec:otherBP}.
Here, we derive results for the flavor-unchanged contributions as
\begin{align}
\delta \mathcal{L}^{\rm box+peng}_{Hq3} &=
- \frac{\alpha}{\pi s_W^2} \left[ \lambda^{kk}_{m'} + \lambda^{ll}_{m'} \right]
(C^{(3)}_{Hq})_{kl} (\mathcal{Q}_L)_{klii}
\notag \\ &~~
\times
\bigg\{
B(x_{m'}) +\Xi_2(x_{m'}) - \frac{1}{16} \left[ 3 -\xi_W -\frac{6\xi_W \ln \xi_W}{\xi_W-1} \right]
\bigg\}
\notag \\ &~~
- \frac{\alpha}{\pi s_W^2} \left[ \lambda^{kk}_{m'} + \lambda^{ll}_{m'} \right] 
(C^{(3)}_{Hq})_{kl} \left[(\mathcal{Q}_L)_{klii} + (\mathcal{Q}_R)_{klii}
\right]
\notag \\ &~~
\times s_W^2
\bigg\{
D(x_{m'})-2 \Xi_2(x_{m'})
\notag \\ &\qquad\quad~~
- \frac{1}{24(\xi_W-1)^3}
\bigg[
\frac{463}{18} - \frac{577}{6}\xi_W +\frac{649}{6}\xi_W^2-\frac{733}{18}\xi_W^3 +3 \xi_W^4 
\notag \\ &\qquad\qquad\qquad\qquad\qquad\quad~~
+ \left(1+3\xi_W -30\xi_W^2 +16\xi_W^3\right)\ln\xi_W 
\bigg]
\bigg\}
\notag \\ &~~
- \frac{\alpha}{\pi s_W^2} \left[ \lambda^{kk}_{m'} + \lambda^{ll}_{m'} \right] 
(C^{(3)}_{Hq})_{kl} (\mathcal{Q}_Z)_{klii}
\notag \\ &~~
\times
\bigg\{
J_3(x_{m'},\mu) + \Xi_2(x_{m'}) - 3 \Xi_1(x_{m'})
- \frac{2}{3} s_W^2
\big[ K_1(x_{m'},\mu)-\Xi_1(x_{m'}) \big]
\notag \\ &\qquad
- \frac{1}{8}
\left[
(3 +2\xi_W) \ln \frac{\mu^2}{M_W^2}
+ \frac{1}{2} (2 + 3\xi_W)
- \frac{2\xi_W +1}{\xi_W -1} \xi_W\ln \xi_W
\right]
\notag \\ &\qquad
+ \frac{1}{48} s_W^2
\bigg[
(9 +7\xi_W) \ln\frac{\mu^2}{M_W^2}
+ \frac{1}{2} \left( 3 +11\xi_W \right)
- (2+7\xi_W) \frac{\xi_W \ln \xi_W}{\xi_W-1}
\bigg]
\bigg\}.
\end{align}
The last term comes from the vertex corrections \eqref{eq:ve2} and those independent of $\lambda^{kk}_{m'}$ and $\lambda^{ll}_{n'}$ as
\begin{align}
\mathcal{L}^{\rm vart}_{Hq3} &= 
- \frac{\alpha}{\pi s_W^2} \lambda^{kk}_{m'}\lambda^{ll}_{n'}
\big[ J_2(x_{m'},x_{n'})+\Xi_1(x_{m'})+\Xi_1(x_{n'})+\Xi_0 (\mu) \big]
(C_{Hq}^{(3)})_{kl}(\mathcal{Q}_Z)_{klii}
\notag \\ &~~
- \frac{\alpha}{\pi s_W^2}(C_{Hq}^{(3)})_{kl}(\mathcal{Q}_L)_{klii}
\notag \\ &~~
\times
\frac{1}{8}
\bigg\{
\left[ (3 +2 \xi_W) -3 s_W^2 (1+\xi_W) \right] \ln \frac{\mu^2}{M_W^2}
\notag \\ &\qquad~~
+ \frac{2 +3\xi_W +s_W^2 (1 -5 \xi_W)}{2}
- \left[ 1+(2-3 s_W^2)\xi_W \right] \frac{\xi_W\ln \xi_W}{\xi_W-1} 
\bigg\}
\notag \\ &~~
- \frac{\alpha}{\pi s_W^2}
(C^{(3)}_{Hq})_{kl} \left[(\mathcal{Q}_L)_{klii}+(\mathcal{Q}_R)_{klii}\right]
\times \frac{1}{12}s_W^2\left[ \ln\frac{\mu^2}{M_W^2}-\ln\xi_W \right].
\end{align}
As a result, we obtain
\begin{align}
\delta \mathcal{L}_{\rm eff} &=
- \frac{\alpha}{\pi s_W^2} \lambda^{kk}_{m'} \lambda^{ll}_{n'}
(C^{(3)}_{Hq})_{kl} (\mathcal{Q}_L)_{klii} 
\notag \\ &~~
\times
\bigg\{
(-1 +2s_W^2)
\big[
J_2(x_{m'},x_{n'}) 
+ J_3(x_{n'},\mu) + J_3(x_{m'},\mu)
\notag \\ &\qquad\qquad\qquad\qquad
- K_0(x_{m'},\mu) - K_0(x_{n'},\mu)
- K_1(x_{m'},\mu) - K_1(x_{n'},\mu)
\big]
\notag \\ &\qquad
+ B(x_{m'}) + B(x_{n'})
+ s_W^2 \big[ D(x_{n'}) + D(x_{m'}) \big]
\notag \\ &\qquad
+ 
\left(\frac{3}{4}+\frac{13}{6}s_W^2 -\frac{7}{2} s_W^4\right) \ln \frac{\mu^2}{M_W^2}
+\frac{5}{8}+s_W^2\left(\frac{113}{108}-\frac{1}{3}s_W^2\right)
\bigg\}
\notag \\ &~~
- \frac{\alpha}{\pi s_W^2} \lambda^{kk}_{m'} \lambda^{ll}_{n'} 
(C^{(3)}_{Hq})_{kl} (\mathcal{Q}_R)_{klii}
\notag \\ &~~
\times
\bigg\{
2 s_W^2
\bigg[
J_2(x_{m'},x_{n'}) 
+ J_3(x_{m'},\mu) + J_3(x_{n'},\mu) 
\notag \\ &\qquad\qquad~~~
- K_0(x_{m'},\mu) - K_0(x_{n'},\mu)
- K_1(x_{m'},\mu) - K_1(x_{n'},\mu)
\bigg]
\notag \\ &\qquad
+ s_W^2 \big[ D(x_{m'})+D(x_{n'}) \big]
\notag \\ &\qquad
+ s_W^2
\left[ 
\left(\frac{13}{6}-\frac{7}{2}s_W^2\right) \ln \frac{\mu^2}{M_W^2} 
+ \frac{113}{108}-\frac{1}{3}s_W^2
\right]
\bigg\}.
\end{align}
Therefore, it is found that the gauge parameters are canceled completely.

\bibliographystyle{utphys28mod}
\bibliography{ref}

\end{document}